\documentstyle[12pt]{article}
\renewcommand{\theequation}{\arabic{section}.\arabic{equation}}

\newcommand{\dfx}{d^4x}

\newcommand{\Amd}{A_{\mu}}
\newcommand{\Amu}{A^{\mu}}

\newcommand{\Dmd}{D_{\mu}}

\newcommand{\Fmnu}{F^{\mu\nu}_+}

\newcommand{\gmnu}{g^{\mu\nu}}

\newcommand{\Sla}[1]{{\not \! \! {#1}}}
\newcommand{\sla}[1]{{\not \! \! \: {#1}}}

\newcommand{\theb}{{\bar \theta}}
\newcommand{\mb}{{\bar m}}
\newcommand{\qd}{q^\dagger}
\newcommand{\psiq}{\psi_q}
\newcommand{\psiqd}{\psi_q^\dagger}
\newcommand{\Hq}{H_q^I}
\newcommand{\Hqd}{H_q^{I\dagger}}
\newcommand{\chiq}{\chi_q^I}
\newcommand{\chiqd}{\chi_q^{I\dagger}}
\newcommand{\Bd}{B^\dagger}

\newcommand{\HB}{H_B^{II}}

\newcommand{\chiB}{\chi_B^{II}}

\newcommand{\Bmnu}{B^{\mu\nu}_+}
\newcommand{\Bmnd}{B_{+\mu\nu}}
\newcommand{\cD}{{\cal D}}
\newcommand{\half}{\frac{1}{2}}
\newcommand{\sigbmnu}{{\bar \sigma}^{\mu\nu}}
\newcommand{\sigbmnd}{{\bar \sigma}_{\mu\nu}}
\newcommand{\sigb}{{\bar \sigma}}
\newcommand{\nonum}{\nonumber}
\newcounter{subeq}

\begin{document}


\begin{titlepage}
\vglue 3cm

\begin{center}
\vglue 0.5cm
{\Large\bf Euler number of Instanton Moduli space
\\ and Seiberg-Witten invariants} 
\vglue 1cm
{\large A.Sako${}^\dagger $,T.Sasaki${}^* $} 
\vglue 0.5cm
{\it ${}^\dagger $Department of Mathematics, Hiroshima University,\\
 Higashi-Hiroshima 739-8526 ,Japan}\\
{\it ${}^*$Department of Physics,
Hokkaido University,Sapporo 060-0810,Japan}
\baselineskip=12pt

\vglue 1cm
{\bf ABSTRACT}
\end{center}
{\rightskip=3pc
 \leftskip=3pc}

\noindent

We show that a partition function of topological 
twisted $N=4$
Yang-Mills theory is given by Seiberg-Witten
invariants on a Riemannian four manifolds under the 
condition that the sum of Euler number and 
signature of the four manifolds vanish.
The partition function is the sum of 
Euler number of instanton moduli space
when it is possible to apply the vanishing theorem.
And we get a relation of Euler number 
labeled by the instanton number $k$ with 
Seiberg-Witten invariants, too.
All calculation in this paper is done
without assuming duality.  
\vspace{5cm}

\begin{flushleft}
\baselineskip=12pt
\hrule
\vspace{0.2cm}
{${}^\dagger $sako@math.sci.hiroshima-u.ac.jp \\
${}^*$ sasaki@particle.sci.hokudai.ac.jp}
\end{flushleft}

\end{titlepage}

\newpage

\section{Introduction}
\hspace*{5mm}
The aim of this paper is to get a relation of 
the partition function of topological
twisted $N=4$ gauge theory with Seiberg-Witten invariants
in four manifolds.\\
The partition function is given by Euler number of 
instanton moduli space in some conditions.
We will show that the Euler number labeled 
by instanton number $k$
is expressed by Seiberg-Witten invariants when the sum 
of Euler number and signature of the base four manifolds
vanishes.
This result gives us the formulas to get the partition function
of the twisted $N=4$ gauge theory by Seiberg-Witten invariants. \\


The partition functions of the $N=4$ Yang-Mills theories on 
some four manifolds 
are calculated by Vafa-Witten with 
topological field theory \cite{vafa-witten} 
\cite{lozano}.
It is an $SL(2,Z)$ modular form.
$SL(2,Z)$ transformation is understood as an extension of 
Montonen-Olive duality \cite{Montonen-Olive}.
So the duality relation is apparent in that partition function.\\

This duality is deeply connected with  
the Hilbert scheme picture of instanton moduli space \cite{mukai}.
But, in general, instanton moduli space has variety compactification 
and the sum of Euler number of any compactified moduli space 
is not necessarily a modular form. 
Actually, in our calculus, the partition function is not modular form
with no contrivance.
On the other side, $N=4$ gauge theory is given by the toroidal compactification of 10-dim $N=1$
gauge theory on a 4-dim manifold.
(Note that "compactification" is used two ways.) 
So the  theory is interpreted as a low energy theory of the Heterotic or TypeI string theory.
Recent developments of string theory show us many evidences of duality
relation in field theory. 
In our case, Vafa shows us one method to link the compactified instanton 
moduli space with the Hilbert scheme \cite{vafa}.
This fact implies that a choice of compactification is understood
in string theory better than field theory.
We discuss the problem of compactification and duality later.\\


For our purpose we use a similar tool to \cite{H-P-P}.
They used the non-abelian monopole theory and related the 
Donaldson invariants to Seiberg-Witten invariants
without using duality \cite{donaldson} \cite{S-W}.
We also calculate the partition function in low energy limit
of cohomological field theory \cite{TFT} and there is no request of S-duality.
This is the most different point from Dijkgraaf-Park-Schroers \cite{DPS}.
They have determined the partition function of $N=4$ supersymmetric
Yang-Mills theory on a K\"ahler surface, using S-duality. 
Their result is given by Seiberg-Witten invariants, too.
So it is interesting to compare our results with theirs.\\


What we do first is to extend the instanton moduli space to 
non-abelian monopole moduli \cite{TQCD} \cite{Ltqcd}.
In usual cohomological field theory, it was done 
in \cite{TQCD}.
Vafa-Witten theory is constructed as a balanced 
topological field theory
(we denote it as BTFT in the following) \cite{BTFT}.
BTFT has no ghost number anomaly, and its partition function
is a sum of Euler number of given zero-section space 
under vanishing theorem.
In the 2nd section, we will construct the non-abelian 
monopole theory as BTFT and 
investigate some character of the theory.
The vanishing theorem is an obstruction to construct the
partition function as the sum of Euler number of the monopole moduli,
and to get a relation with Vafa-Witten theory.
We do not study this case closer in this paper. \\
In the 3rd section, we get the formulas
between the partition function of 
a twisted $N=4$ Yang-Mills theory and Seiberg-Witten
invariants. To get them, we break the balance of topological charge.
The tools in this paper were used in getting a relation of 
Donaldson invariants and Seiberg-Witten invariants \cite{H-P-P}.
We use a model which has a gauge multiplet that is balanced
and a hypermultiplet that is not balanced.
We call the model unbalanced topological QCD.
Vacuum expectation value of an observable is calculated
and the relation between Euler number of instanton moduli
space and Seiberg-Witten invariants is obtained if vanishing theorem
is applicable and the sum of Euler number and signature of the four manifolds
vanishes. 
The comparison with \cite{vafa-witten,DPS} is also made in this section.
At the last section, we discuss some remaining problems and 
the possibility of extension.\\

 
\section{Balanced Topological QCD}
\label{sec:2}
\setcounter{equation}{0}
\hspace*{5mm}
In this section,  we 
construct
a Balanced Topological QCD (BTQCD),
which is a twisted $N=4 $ Yang-Mills theory 
coupled with massive hypermultiplets in the
fundamental representation
\cite{DPS,TQCD,H-P-P}.

\subsection{Balanced Topological QCD}
\label{sec:2-1}
\hspace*{5mm}
Let $X $ be a compact Riemannian four manifold 
and $E $ be an $SU(2) $-bundle over  $X $.
The bundle  $E $ is classified by the instanton number

\begin{equation}
\label{2.1}
k=\frac{1}{8\pi^2}\int_X Tr F\wedge F,
\end{equation}
where $Tr $ is the trace in the fundamental representation 
of $SU(2) $ and $F\in\Omega^2_X({\cal G}_E) $ is
the adjoint valued curvature 2-form on  $X $.
We denote the group of gauge transformations by
${\cal G} $, i.e. elements of ${\cal G} $ are 
 sections of $P $,
where $P $ is the associated principal $SU(2) $-bundle 
over $X $.
We pick a $spin^c $ structure $c $ on  $X $ and
consider the associated $spin^c $ bundle $W^\pm_c $.
Let $\cal A $ be the space of all connections on $P $ and
$\Gamma(W^+_c\otimes E)(\Gamma(W^-_c\otimes E)) $ 
the space of the sections of the $spin^c $ bundle twisted by
the vector bundle $E $.
After twisting, the complex boson in the hypermultiplet
becomes a section of $\Gamma(W^+_c\otimes E)(\Gamma(W^-_c\otimes E))$;
\[
q\in \Gamma(W^+_c\otimes E),~\qd\in\Gamma({\bar W}^+_c\otimes{\tilde E}),
\]
\begin{equation}
  \label{2.2}
B\in \Gamma(W^-_c\otimes E), ~\Bd\in\Gamma({\bar W}^-_c\otimes{\tilde E}),
\end{equation}
where ${\tilde E}$ denotes the vector bundle conjugate to $E $.
The $spin^c $ Dirac operator 
\begin{equation}
  \label{2.3}
\sigma^\mu D_\mu :\Gamma(W^+_c\otimes E)\to \Gamma(W^-_c\otimes E),
\end{equation}
is the Dirac operator for the $spin^c $ bundle twisted by  $E $.
We will sometimes denote $\sigma^\mu D_\mu$ by
${\Sla D} $ or ${\Sla D}^E_c $.

Throughout this paper, we restrict our attention to the case that
the gauge group is $SU(2) $ and the theory is coupled with
hypermultiplets 
in
the fundamental representation.

\paragraph{algebra of BTQCD}~~  \\

In this paragraph, the algebra of BTQCD is given.

We introduce two global supercharges $Q_\pm $
carrying an additive quantum number (ghost number) $U=\pm 1 $.
When they act on fields 
in the adjoint representation,
they satisfy the following commutation relations:
\begin{equation}
  \label{2.4}
Q^2_+=\delta^g_\theta,~\{Q_+,Q_- \}=-\delta^g_c,~Q^2_-=-\delta^g_\theb,
\end{equation}
where $\delta^g_\theta$ denotes the gauge transformation generated by
adjoint scalar field $\theta\in \Omega^0_X({\cal G}_E) $
and we adopt 
$\delta^g_\theta \Amd=\Dmd\theta,~
\delta^g_\theta\Bmnd=i[\Bmnd,\theta],~
\delta^g_\theta c=i[c,\theta]$.
When they act on fields
in the fundamental representation,
they satisfy the following commutation relations:
\begin{equation}
  \label{2.5}
Q^2_+=-\delta^g_\theta,~\{Q_+,Q_- \}=\delta^g_c,~Q^2_-=\delta^g_\theb,
\end{equation}
where we also introduce $U(1)$ global transformation
generated by $m\in i R $
and we adopt 
$\delta^g_\theta q=(i\theta+m)q,~\delta^g_\theta \qd=\qd(-i\theta-m),~
\delta^g_\theta B=(i\theta+m)B,~ \delta^g_\theta \Bd=\Bd(-i\theta-m)$.
The relative sign difference between (\ref {2.4}) and (\ref {2.5})
is simply the difference of representations.
A simple explanation is the following.
One can construct a field $J^a $ in the adjoint representation
with a pair of fields $q,\qd $ in the fundamental representation,
\begin{equation}
J^a\equiv \qd T^a q.
\end{equation}
Using above transformations,
one can check (\ref{2.4}) follows from (\ref{2.5}),
\begin{eqnarray}
Q_+^2 J^a &=& Q_+^2 (\qd T^a q)
\nonum
\\
&=&
(-\delta^g_\theta \qd)T^a q+\qd T^a (-\delta^g_\theta q)
\nonum
\\
&=&
i[\qd T q ,\theta]^a
\nonum
\\
&= &\delta^g_\theta J^a.
\end{eqnarray}
Note that the relative sign difference between (\ref {2.4}) and (\ref {2.5})
is consistent with this derivation.
The recipe for giving mass to 
fields
in the fundamental representation
by global symmetry
is considered by Hyun-Park-Park(H-P-P)\cite{H-P-P}.

We define $\delta_\pm $ transformations $\delta_\pm\equiv[Q_\pm,* \} $.
$\delta_\pm $ transformations are given in
Appendix A. 
See also the references \cite{BTFT,H-P-P}.

\paragraph{action of BTQCD}~ \\

Using above fields and transformations,
we define the action of BTQCD as

\begin{equation}
  \label{2.9}
h^2 S=\int \sqrt g {\cal L},
\end{equation}
where
\begin{equation}
  \label{2.10}
{\cal L}=\delta_+\delta_-{\cal F}.
\end{equation}
${\cal F} $ is described with fields in the previous paragraph
and has ghost number $0 $.
The general recipe for
constructing a balanced topological field theory
is given by Moore et.al\cite{BTFT}.

${\cal F} $ is explicitly given by,
\begin{eqnarray}
{\cal F}&=&
(B^{\mu\nu a}_+s_{+\mu\nu}^a)
-(\chi^{I\mu\nu a}_+\psi_{B\mu\nu}^a)
-(\chi^{IIa}_{B\mu}\psi^{\mu a})
+(-i\frac{1}{3}{\Bmnu}^a[B_{+\mu\rho},B_{+\nu\sigma}]^ag^{\rho\sigma})
\nonum
\\
& &
+(B^{\dagger\alpha}s_\alpha)
-(\chi^{I\dagger\alpha}_q\psi_{B\alpha})
-(\chi^{II\dagger}_{B\dot{\alpha}}\psi^{\dot{\alpha}}_q)
\nonum
\\
& &
+(s^{\dagger\alpha}B_\alpha)
+(\psi^{\dagger\alpha}_B\chi^I_{q\alpha})
+(\psi^\dagger_{q\dot{\alpha}}\chi^{II\dot{\alpha}}_B)
\nonum
\\
& &
+(\xi^a\eta^a),
\label{2.11}
 \end{eqnarray}
where
\begin{equation}
  \label{2.13}
s_+^{\mu\nu}=\Fmnu+\qd\sigbmnu q
\end{equation}
\begin{equation}
  \label{2.14}
s_\alpha=({\Sla D}q)_\alpha.
\end{equation}

Finally, full lagrangian is given by

\begin{eqnarray}
{\cal L}^{full} = \delta_+\delta_-{\cal F}.
\label{2.15}
\end{eqnarray}
Explicit expression of this lagrangian
is given in appendix A.
This lagrangian (\ref{full}) is different from \cite{vafa-witten}
in matter fields ($q,B $ etc.)
and also different from H-P-P\cite{H-P-P} in dual fields ($\Bmnu,c,B$ etc.).
But due to its construction,
it is balanced.

\subsection{Fixed Point}
\label{sec:2-2}
\hspace*{5mm}
In this subsection, we study 
the nature of the
action given in subsection \ref{sec:2-1}.
Here in particular we investigate
the fixed points and vanishing theorem\cite{vafa-witten}.

\paragraph{Fixed Point}~ \\

To check the nature of lagrangian, 
we decompose the bosonic part of lagrangian (\ref{full})

\begin{equation}
  \label{2.17}
{\cal L}^{full}_{boson}={\cal L}^{eq}_{boson}+{\cal L}^{pro}_{boson},
\end{equation}
where
\begin{eqnarray}
  {\cal L}^{eq}_{boson}&=&
-H^{I\mu\nu a}_+\{
H^{Ia}_{+\mu\nu}-(
s_{+\mu\nu}^a-i[B_{+\mu\rho},B_{+\nu\sigma}]^ag^{\rho\sigma}
-i[B_{+\mu\nu},c]^a
)
\}
\nonum
\\
& &
-H^{II\rho a}_B\{
H^{IIa}_{B\rho}-(
-2D^\mu B_{+\mu\rho}^a
+iB^\dagger \sigma_\rho T^aq-iq^\dagger{\bar{\sigma}_\rho}T^aB
-D_\rho c^a
)
\}
\nonum
\\
& &
-H^{I\dagger\alpha}_q\{
H^I_{q\alpha}-(
s_\alpha+icB_\alpha+m_cB_\alpha
)
\}+(h.c.)
\nonum
\\
& &
-H^{II\dagger}_{B\dot{\alpha}}\{
H^{II\dot{\alpha}}_B-(
-(\Sla{D}B)^{\dot{\alpha}}
+({\bar{\sigma}}^{\mu\nu} B_{+\mu\nu}   q)^{\dot{\alpha}}
+icq^{\dot{\alpha}}+m_cq^{\dot{\alpha}}
)
\}+(h.c.)
\nonum
\\
& &
\label{2.18}
\end{eqnarray}
and
\begin{eqnarray}
   {\cal L}^{pro}_{boson}&=&
-\{
[\theta,{\bar{\theta}}]^a[{\bar{\theta}},\theta]^a
-[c,\theta]^a[c,{\bar{\theta}}]^a
+[B^{\mu\nu}_+,{\bar{\theta}}]^a[B_{+\mu\nu},\theta]^a
\}+D_\mu{\bar{\theta}}^aD^\mu\theta^a
\nonum
\\
& &
+(
-iq^\dagger{\bar{\theta}}-q^\dagger{\bar{m}}
)
(i\theta q+mq)
+(
-iq^\dagger\theta-q^\dagger m
)
(i{\bar{\theta}}q+{\bar{m}}q)
\nonum
\\
& &
+(
-iB^\dagger{\bar{\theta}}-B^\dagger{\bar{m}}
)
(i\theta B+mB)
+(
-iB^\dagger\theta-B^\dagger m
)
(i{\bar{\theta}}B+{\bar{m}}B).
\label{2.19}
\end{eqnarray}
${\cal L}^{eq}_{boson} $
is defining the moduli space that we want to consider and
${\cal L}^{pro}_{boson} $
is induced for the projection to gauge normal direction.
(\ref{2.18})  lagrangian is rewritten as

\begin{eqnarray}
{\cal L}^{eq}_{boson}&=&
H~~ square~~ terms
\nonum
\\
& &
+\frac{1}{4}(s_{+\mu\nu}^a-i[B_{+\mu\rho},B_{+\nu\sigma}]^ag^{\rho\sigma}
)^2
-\frac{1}{4}([B_{+\mu\nu},c]^a)^2
\nonum
\\
& &+\frac{1}{4}(
-2D^\mu B_{+\mu\rho}^a
+iB^\dagger \sigma_\rho T^aq-iq^\dagger{\bar{\sigma}_\rho}T^aB)^2
+\frac{1}{4}(D_\rho c^a)^2
\nonum
\\
& &
+\frac{1}{2}\vert s\vert^2
+\frac{1}{2}\vert icB+m_c B\vert^2
\nonum
\\
& &
+\frac{1}{2}\vert
-(\Sla{D}^\dagger B)^{\dot{\alpha}}
+({\bar{\sigma}}^{\mu\nu} B_{+\mu\nu}   q)^{\dot{\alpha}}\vert^2
+\frac{1}{2}\vert icq+m_c q\vert^2.
\label{2.21}
\end{eqnarray}
Thus we have the following fixed point equations

\[
F_{+\mu\nu}+\qd \sigbmnd q-i[B_{+\mu\rho},B_{+\nu\sigma}]g^{\rho\sigma}
=0
\]
\[
-2D_\mu \Bmnu+i\Bd \sigma^\nu q-i\qd {\bar \sigma}^\nu B=0
\]
\[
s={\Sla D}q=0
\]
\[
-{\Sla D}^\dagger B+\sigbmnd \Bmnu q=0
\]
\[
D_\nu \theta=D_\nu c=D_\nu \theb=0
\]
\[
[\theta,\theb]=[c,\theta]=[c,\theb]=[\Bmnu,\theta]=[\Bmnu,\theb]=[\Bmnu,c]=0
\]
\[
(i\theta+m)q=(i\theb+\mb)q=(ic+m_c)q=0
\]
\[
\qd(-i\theta-m)=\qd(-i\theb-\mb)=\qd(-ic-m_c)=0
\]
\[
(i\theta+m)B=(i\theb+\mb)B=(ic+m_c)B=0
\]
\begin{equation}
  \label{2.22}
\Bd(-i\theta-m)=\Bd(-i\theb-\mb)=\Bd(-ic-m_c)=0.
\end{equation}
If hypermultiplet fields are set to zero ($q=\qd=B=\Bd=0 $),
then above equations are Vafa-Witten equations
\cite{vafa-witten}\cite{DPS}.
Thus we call above equations Extended Vafa-Witten equations.

\paragraph{problem}~ \\

In the previous paragraph, we have obtained 
fixed point equations of BTQCD.
The equations for fermionic zero-modes are just the linearization
of the fixed point equation and the condition that they are orthogonal
to gauge orbits.
Due to the balanced structure each fermionic zero-mode
has a partner with the opposite $U $-number.
Thus there is no ghost-number anomaly and the partition function is
well defined,
i.e.there is no need to insert observables.
We want to compute the partition function of BTQCD.
According to Vafa-Witten if an appropriate vanishing theorem holds,  
the partition function becomes the sum of 
Euler number 
of moduli space which we want to calculate.
Roughly speaking, vanishing theorem is understood 
as the condition that dual fields ($\Bmnd,c,B,\Bd $ etc.)
are to be zero and the dimensions of their moduli space 
become zero,
when we choose an appropriate metric\cite{vafa-witten}.
But we could not verify that vanishing theorem holds in this model.
To compare the result of this section to that of 
the next section,
we  give  the only result to compute 
the partition function of BTQCD on 
the condition that
vanishing theorem holds.

\subsection{result}
\label{sec:2-3}
\hspace*{5mm}
In this subsection, we give the result of computing
 the path integral of BTQCD.
We define partition function of BTQCD as

\begin{equation}
  \label{2.23}
Z=\frac{1}{Vol{\cal G}(2\pi)^\Omega}
\int
\cD W\cD\psi_W\cD Q^\dagger\cD\psi^\dagger_Q \cD Q \cD \psi_Q
e^{-S},
\end{equation}
where

\[
W=A_\mu,\Bmnu,H^\mu_B,H^{\mu\nu}_+,\theta,c,\theb
\]
\[
\psi_W=\psi_\mu,\psi_B^{\mu\nu},\chi^{II\mu}_B,\chi^{I\mu\nu}_+,\xi,\eta
\]
\[
Q=q,B,H^I_q,H^{II}_B
\]
\[
\psi_Q=\psi_q,\psi_B,\chi^I_q,\chi^{II}_B
\]
\begin{equation}
  \label{2.24}
\Omega=\dim of~H's.
\end{equation}
Here we denote auxiliary fields as $H^\mu_B, H^{\mu\nu}_+, 
H^I_q,H^{II}_B$,
and we call auxiliary fields for $Y$ as $H's~of~Y$
in the following. $\dim of~H's $ is a number of the auxiliary fields.

After path-integrations of transverse part 
we get the partition function as the 
sum of two branches, according to the methods of the next section,
\begin{equation}
  \label{2.25}
Z=Z^{V-W}+Z^{B-U(1)S-W}.
\end{equation}
$Z^{V-W} $ is a contribution from branch 1)
(gauge symmetry is unbroken),
and corresponds to Vafa-Witten partition function. 
$Z^{B-U(1)S-W}$
is a contribution from branch 2)
(gauge symmetry is broken to $U(1)$),
and corresponds to balanced $U(1)$ monopole theory.
The fixed point equations of the balanced $U(1)$ monopole theory
are

\[
F_{+\mu\nu}^3+\half{\qd}_1 \sigbmnd q_1
=0
\]
\[
-2\nabla_\mu {\Bmnu}^3
+i\half{\Bd}_1 \sigma^\nu q_1-i\half{\qd}_1 {\bar \sigma}^\nu B_1=0
\]
\[
{\Sla D}^3 q_1=0
\]
\begin{equation}
-{\Sla D}^{\dagger 3} B_1+\half\sigbmnd {\Bmnu}^3 q_1=0.
\label{2.27}
\end{equation}
Where $F_{+\mu\nu}^3$ is a curvature of $U(1)$ left symmetry
after breaking $SU(2)$ and the labels of $q_1$ are $B_1$
are the ones of color.
Since we do not know the vanishing theorem 
for dual fields (${\Bmnd}^3,B_1,{\Bd}_1 $) 
from (\ref{2.27}), we stop to investigate this model
further more in this paper.

\section{Unbalanced Topological QCD}
\label{sec:3}
\setcounter{equation}{0}
\hspace*{5mm}
In this section, we compute a correlation function of
an appropriate BRS exact operator in Unbalanced Topological QCD.
As a result, we can describe Euler number of
instanton moduli space with Seiberg-Witten invariants.
We have a similar but not the same expression to Dijkgraaf et.al\cite{DPS},
because we treat the different theory from theirs.
We discuss this point at the end of this section.

\subsection{Unbalanced Topological QCD}
\label{sec:3-1}
\hspace*{5mm}
Here we construct Unbalanced Topological QCD,
which is a twisted $N=4$ Yang-Mills theory coupled with only one massive
hypermultiplet in the fundamental representation
(we denote it as UBTQCD in the following).
Alternatively one get a UBTQCD,
when one set one massive hypermultiplet $(B,\psi_B,\chiB,\HB) $ 
of BTQCD 
in the previous section
to zero
(we call this process breaking balanced structure).

\paragraph{algebra of UBTQCD}~ \\

The algebra of UBTQCD is given as a part of the 
BTQCD algebra.
Contrary to the previous section, we only consider
the global supercharge $Q_+ $.
When it acts on adjoint (fundamental) fields,
it satisfies the following commutation relation:
\begin{equation}
Q_+^2=\delta^g_\theta(-\delta^g_\theta).
\end{equation}
We adopt the same $\delta_+$ transformations as the previous section and appendix A of
(\ref {2.7.1})$\sim $(\ref{2.8.2}).

\paragraph{action of UBTQCD}~ \\

We define the action of UBTQCD as

\begin{equation}
h^2 S=\int \dfx \sqrt g {\cal L}
\end{equation}
where
\begin{equation}
{\cal L}=\delta_+\Psi .
\end{equation}

We explicitly give $\Psi $ as;

\begin{eqnarray}
\Psi&=&
-\chi^{I\mu\nu a}_+\{
H^{Ia}_{+\mu\nu}-(
s_{+\mu\nu}^a-i[B_{+\mu\rho},B_{+\nu\sigma}]^ag^{\rho\sigma} 
-i[B_{+\mu\nu},c]^a
)
\}
\nonum
\\
& &
-\chi^{II\rho a}\{
H^{IIa}_{B\rho}-(
-2D^\mu B_{+\mu\rho}^a
-D_\rho c^a
)
\}\nonum
\\
& &
-\chi^{I\dagger\alpha}_q\{
H^I_{q\alpha}-
s_\alpha
\}\nonum
\\
& &
-\{
H^{I\dagger\alpha}_q-
s^{\dagger\alpha}
\}\chi^I_{q\alpha}
\nonum
\\
& &
+\{ i[\theta,{\bar{\theta}}]^a\eta^a
-i\xi^a[c,{\bar{\theta}}]^a \}
+i[B^{\mu\nu}_+,{\bar{\theta}}]^a\psi_{B\mu\nu}^a
+D_\mu{\bar{\theta}}^a\psi^{\mu a}
\nonum
\\
& &
-(
-iq^\dagger_{\dot{\alpha}}{\bar{\theta}}-{\bar{m}}q^\dagger_{\dot{\alpha}}
)\psi^{\dot{\alpha}}_q
-\psi^\dagger_{q\dot{\alpha}}(
i{\bar{\theta}}q^{\dot{\alpha}}+{\bar{m}}q^{\dot{\alpha}}
),
\end{eqnarray}
where 

\begin{equation}
s^{\mu\nu a}_+={\Fmnu}^a+\qd\sigbmnu T^a q
\end{equation}
\begin{equation}
s^\alpha=({\Sla D}q)^\alpha .
\end{equation}

Finally the full lagrangian is given by

\begin{eqnarray}
{\cal L}^{full}&=&\delta_+\Psi
\\
&=&
-H^{I\mu\nu a}_+\{
H^{Ia}_{+\mu\nu}-(
s_{+\mu\nu}^a-i[B_{+\mu\rho},B_{+\nu\sigma}]^ag^{\rho\sigma}
-i[B_{+\mu\nu},c]^a
)
\}\nonum
\\
& &
-\chi^{I\mu\nu a}_+\{
-i[\chi^I_{+\mu\nu},\theta]^a
+2D_\mu\psi_\nu^a
+\psi^\dagger_q{\bar{\sigma}}_{\mu\nu}T^aq
+q^\dagger{\bar{\sigma}}_{\mu\nu}T^a\psi_q
-2i[B_{+\mu\rho},\psi_{B\nu\sigma}]^ag^{\rho\sigma}
\nonum
\\
& &
-i[\psi_{B\mu\nu},c]^a
-i[B_{+\mu\nu},\xi]^a
\}\nonum
\\
& &
-H^{II\rho a}_B\{
H^{IIa}_{B\rho}-(
-2D^\mu B_{+\mu\rho}^a
-D_\rho c^a
)
\}\nonum
\\
& &
-\chi^{II\rho a}_B\{
-i[\chi^{II}_{B\rho},\theta]^a
-2D^\mu\psi_{B\mu\rho}^a
-2i[\psi^\mu,B_{+\mu\rho}]^a
-D_\rho\xi^a
-i[\psi_\rho,c]^a
\}\nonum
\\
& &
-H^{I\dagger\alpha}_q\{
H^I_{q\alpha}-
s_\alpha
\}\nonum
\\
& &
-\chi^{I\dagger}_q\{
{\Sla D}\psi_q
+\sigma_\rho i\psi^\rho q
\}\nonum
\\
& &+(h.c.~~ above~~ two~~ lines)
\nonum
\\
& &
-\{
[\theta,{\bar{\theta}}]^a[{\bar{\theta}},\theta]^a
-[c,\theta]^a[c,{\bar{\theta}}]^a
+[B^{\mu\nu}_+,{\bar{\theta}}]^a[B_{+\mu\nu},\theta]^a
\}+D_\mu{\bar{\theta}}^a D^\mu\theta^a
\nonum
\\
& &
i[\theta,\eta]^a\eta^a
+i\xi^a[\xi,{\bar{\theta}}]^a
+i\xi^a[c,\eta]^a
+i[\psi^{\mu\nu}_B,{\bar{\theta}}]^a\psi_{B\mu\nu}^a
+i[B^{\mu\nu}_+,\eta]^a\psi_{B\mu\nu}^a
+D_\mu\eta^a\psi^{\mu a}
+i[\psi_\mu,{\bar{\theta}}]^a\psi^{\mu a}
\nonum
\\
& &
+(
-iq^\dagger{\bar{\theta}}-q^\dagger{\bar{m}}
)
(i\theta q+mq)
+(
-iq^\dagger\theta-q^\dagger m
)
(i{\bar{\theta}}q+{\bar{m}}q)
\nonum
\\
& &
+2\psi^\dagger_q(i{\bar{\theta}}+{\bar{m}})\psi_q
-2\chi^{I\dagger}_q(i\theta+m)\chi^I_q
-(
-iq^\dagger\eta-q^\dagger\eta_m
)\psi_q
+\psi^\dagger_q(
i\eta q+\eta_m q
).\label{3.10}
\end{eqnarray}
Notice that lagrangian  (\ref{3.10}) 
is given by lagrangian  (\ref{full}) of previous section
if $(B,\psi_B,\HB,\chiB) $ is set to zero.

\subsection{Fixed Point}
\label{sec:3-2}
\hspace*{5mm}
In this subsection, we study 
the nature of the
action given in subsection \ref {sec:3-1}.
Here in particular we investigate the fixed points
and some observable to insert. 

\paragraph{Fixed Point}~\\

To check the nature of lagrangian,
we decompose the bosonic part of lagrangian (\ref{3.10})

\begin{equation}
{\cal L}^{full}_{boson}={\cal L}^{eq}_{boson}+{\cal L}^{pro}_{boson},
\end{equation}
where

\begin{eqnarray}
  {\cal L}^{eq}_{boson}&=&
-H^{I\mu\nu a}_+\{
H^{I a}_{+\mu\nu}-(
s_{+\mu\nu}^a-i[B_{+\mu\rho},B_{+\nu\sigma}]^ag^{\rho\sigma}
-i[B_{+\mu\nu},c]^a
)
\}\nonum
\\
& &
-H^{II\rho a}_B\{
H^{II a}_{B\rho}-(
-2D^\mu B_{+\mu\rho}^a
-D_\rho c^a
)
\}\nonum
\\
& &
-H^{I\dagger\alpha}_q\{
H^I_{q\alpha}-
s_\alpha
\}+(h.c.)\label{3.12}
\end{eqnarray}
and

\begin{eqnarray}
   {\cal L}^{pro}_{boson}&=&
-\{
[\theta,{\bar{\theta}}]^a[{\bar{\theta}},\theta]^a
-[c,\theta]^a[c,{\bar{\theta}}]^a
+[B^{\mu\nu}_+,{\bar{\theta}}]^a[B_{+\mu\nu},\theta]^a
\}+D_\mu{\bar{\theta}}^aD^\mu\theta^a
\nonum
\\
& &
+(
-iq^\dagger{\bar{\theta}}-q^\dagger{\bar{m}}
)
(i\theta q+mq)
+(
-iq^\dagger\theta-q^\dagger m
)
(i{\bar{\theta}}q+{\bar{m}}q).
\end{eqnarray}
${\cal L}^{eq}_{boson} $
is defining the moduli space 
that we want to consider here and
${\cal L}^{pro}_{boson} $
is induced for the projection to gauge normal direction.
(\ref{3.12})  lagrangian is transformed into

\begin{eqnarray}
{\cal L}^{eq}_{boson}&=&
-\{
H^{Ia}_{+\mu\nu}-\frac{1}{2}(
s_{+\mu\nu}^a-i[B_{+\mu\rho},B_{+\nu\sigma}]^ag^{\rho\sigma}
-i[B_{+\mu\nu},c]^a
)
\}^2\nonum
\\
& &-\{
H^{IIa}_{B\rho}-\frac{1}{2}(
-2D^\mu B_{+\mu\rho}^a
-D_\rho c^a
)
\}^2\nonum
\\
& &
-2\vert
H^I_{q\alpha}-\half
s_\alpha
\vert^2\nonum
\\
& &
+\frac{1}{4}(
s_{+\mu\nu}^a
-i[B_{+\mu\rho},B_{+\nu\sigma}]^ag^{\rho\sigma}
-i[B_{+\mu\nu},c]^a
)^2\nonum
\\
& &
+\frac{1}{4}(
-2D^\mu B_{+\mu\rho}^a
-D_\rho c^a
)^2\nonum
\\
& &
+\frac{1}{2}\vert s_\alpha\vert^2.
\end{eqnarray}

Thus we have the following fixed point equations

\[
F_{+\mu\nu}+\qd \sigbmnd q-i[B_{+\mu\rho},B_{+\nu\sigma}]g^{\rho\sigma}
-i[B_{+\mu\nu},c]
=0
\]
\[
-2D_\mu \Bmnu-D^\nu c
=0
\]
\[
s={\Sla D}q=0
\]
\[
D_\nu \theta=D_\nu \theb=0
\]
\[
[\theta,\theb]=[c,\theta]=[c,\theb]=[\Bmnu,\theta]=[\Bmnu,\theb]=0
\]
\[
(i\theta+m)q=(i\theb+\mb)q=0
\]
\begin{equation}
\label{3.13}
\qd(-i\theta-m)=\qd(-i\theb-\mb)=0.
\end{equation}

\paragraph{problem}~ \\

In the previous paragraph, we have obtained the 
fixed point equations of
UBTQCD. In the same way as the previous section, the equations
for fermionic zero-modes are just the linearization of the fixed
point equations and the conditions that they are orthogonal to 
gauge orbits.
Compared with the previous section,
UBTQCD does not have balanced structure.
In particular the hypermultiplet does not have balanced structure,
while adjoint representation fields still have balanced structure.
The partition function of unbalanced theory becomes
zero due to its ghost number anomaly when the moduli space dimension
of mater field is non-zero.
Thus to get an well-defined path integral,
we have to insert some observable.
One can think
an observable
\begin{equation}
I=\int \dfx( 
\qd(i\theta +m)q+\psiqd\psiq
).
\end{equation}
Note that this observable itself is BRS exact, i.e.
\begin{equation}
I=\delta_+\half \int \dfx(
-\psiqd q +\qd \psiq
).
\end{equation}
Thus the expectation value of $ I $ is zero 
according to Ward-Takahashi identity,
and the expectation value of $e^ I $ 
becomes zero when this theory 
has ghost number anomaly.
However as we will see,
we obtain non-trivial results.

\subsection{branch}
\label{sec:3-3}
\hspace*{5mm}
In this subsection, we will show that the 
fixed point equations 
are decomposed to two branches.
We take a similar treatment for (\ref{3.13})
to \cite{H-P-P}.

Equations
\begin{equation}
D_\nu \theta=D_\nu \theb=0,
[\theta,\theb]=0
\end{equation}
imply  that $\theta,\theb$ can be diagonalized in the fixed points.
If connection $A_\mu $ are irreducible,  $\theta,\theb$ should be zero
(the gauge symmetry is unbroken ).
If connection $A_\mu $ are reducible,
$\theta,\theb$ can be non-zero
(the gauge symmetry is broken down to $U(1) $).
When these solutions are applied to
\[
(i\theta+m)q=(i\theb+\mb)q=0
\]
\begin{equation}
\qd(-i\theta-m)=\qd(-i\theb-\mb)=0,
\label{3.22}
\end{equation}
we have two branches;\\
branch 1) $\theta=\theb=0$ and $q=\qd=0 $\\
or\\
branch 2) $\theta=\theta^3T^3\ne 0, \theb=\theb^3 T^3\ne 0$ 
and $q\ne 0,\qd\ne 0 $. \\
Note that in the branch 2) we choose unbroken $U(1) $ as
$T^3 $ direction without a loss of generality.

\paragraph{branch 1)}
 $\theta=\theb=0$ and $q=\qd=0 $, i.e.
the gauge symmetry is unbroken.
Remaining fixed point equations are

\[
\Fmnu-i[B_{+\mu\rho},B_{+\nu\sigma}]g^{\rho\sigma}  =0,
-2D_\mu \Bmnu=0,\Dmd c=0
\]
\begin{equation}
[\Bmnu,c]=0.
\end{equation}
Here one may apply the same condition as Vafa-Witten\cite{vafa-witten}
to induce the vanishing theorem,
and get the moduli space of
\begin{equation}
  \Fmnu=0.
\end{equation}

\paragraph{branch 2)}
$\theta=\theta^3T^3\ne 0, \theb=\theb^3 T^3\ne 0$ 
and $q\ne 0,\qd\ne 0 $,
i.e.the gauge symmetry is broken to $U(1)$.
Thus the bundle $E$ splits into line bundles,
$E=L\oplus L^{-1} $ with $L\cdot L=-k $.
Then equations (\ref{3.22})  are
\[
(i\theta^3 T^3+m)q=\left(
  \begin{array}[c]{cc}
\frac{i}{2}\theta^3+m&0\\
0&-\frac{i}{2}\theta^3+m
  \end{array}
\right)
\left(
  \begin{array}[c]{c}
q_1\\
q_2
  \end{array}
\right)=0
\]
\[
(i\theb^3 T^3+\mb)q=\left(
  \begin{array}[c]{cc}
\frac{i}{2}\theb^3+\mb&0\\
0&-\frac{i}{2}\theb^3+\mb
  \end{array}
\right)
\left(
  \begin{array}[c]{c}
q_1\\
q_2
  \end{array}
\right)=0
\]
\[
\qd(-i\theta^3 T^3-m)=
\left(
  \begin{array}[c]{cc}
\qd_1&
\qd_2
  \end{array}
\right)
\left(
  \begin{array}[c]{cc}
-\frac{i}{2}\theta^3-m&0\\
0&\frac{i}{2}\theta^3+m
  \end{array}
\right)
=0
\]

\begin{equation}
\qd(-i\theb^3 T^3-\mb)=
\left(
  \begin{array}[c]{cc}
\qd_1&
\qd_2
  \end{array}
\right)
\left(
  \begin{array}[c]{cc}
-\frac{i}{2}\theb^3-\mb&0\\
0&\frac{i}{2}\theb^3+\mb
  \end{array}
\right)
=0.
\end{equation}
Thus the only non-trivial solutions for $q $ are either
\begin{equation}
q=\left(
\begin{array}[c]{c}
q_1\\
0
\end{array}
\right),
\qd=
\left(
\begin{array}[c]{cc}
\qd_1&
0
\end{array}
\right)
and~
\frac{i}{2}\theta^3+m=\frac{i}{2}\theb^3+\mb=0
\end{equation}
or
\begin{equation}
q=\left(
\begin{array}[c]{c}
0\\
q_2
\end{array}
\right),
\qd=
\left(
\begin{array}[c]{cc}
0&
\qd_2
\end{array}
\right)
and~
-\frac{i}{2}\theta^3+m=-\frac{i}{2}\theb^3+\mb=0.
\end{equation}
Throughout this paper we pick
the non-trivial solutions for $q $
as $q_1\ne 0~ and~\theta^3=2im $.
In this branch the equations
\begin{equation}
[c,\theta]=[c,\theb]=[\Bmnu,\theta]=[\Bmnu,\theb]=0
\end{equation}
imply that non-zero solutions of $\Bmnu,c $ 
have the same direction $T^3$ as $\theta$.
Finally we get remaining equations
\[
F^3_{+\mu\nu}+\half \qd_1  \sigbmnd q_1=0
\]
\[
-2\nabla^\mu \Bmnd^3 =\partial^\mu c^3=0
\]
\begin{equation}
\sigma^\mu {\cal D}_\mu q_1=0,
\label{3.24}
\end{equation}
where $\nabla^\mu $ is the covariant derivative
in respect of Levi-Civita connection of background metric $\gmnu $.
Here we reinterpret $U(1)\otimes U(1) $ 
(gauge $U(1)$ and $spin^c~U(1) $)
as a new $U(1) $ ($spin^{c^\prime}~U(1) $),
or alternately we redefine $W^+_c\otimes \zeta =W_{c^\prime}^+ $
as a different $spin^c $ structure
$c^\prime=c+2\zeta $, i.e., $\det(W^+_c\otimes \zeta)=L_c\otimes\zeta^2$.
As a result, (\ref{3.24}) can be interpreted as 
a perturbed Seiberg-witten monopole equation for the $spin^c $
structure $c^\prime $ as well as H-P-P\cite{H-P-P}
and $B_+,c $ equations for $T^3$ direction.

\subsection{Gaussian integral}
\label{sec:3-4}
\hspace*{5mm}
In this subsection we compute the path integral of UBTQCD.
According to Appendix, we could evaluate the exact path integral
of this theory. In this subsection, we only denote 
the diagonal part of the big matrix (see Appendix)
to read the right contribution easily.
As we have already mentioned in subsection \ref{sec:3-2},
we have to insert some observable of fundamental fields
to get an well-defined path integral.
Thus we define expectation value of $e^I$ as

\begin{equation}
<e^I>_{m,c,k}=\frac{1}{Vol{\cal G}(2\pi)^\Omega}
\int
\cD W\cD\psi_W\cD Q^\dagger\cD\psi^\dagger_Q \cD Q \cD \psi_Q
e^{-S+I},\label{3.29}
\end{equation}
where
\[
W=A_\mu,\Bmnu,H^\mu_B,H^{\mu\nu}_+,\theta,c,\theb
\]
\[
\psi_W=\psi_\mu,\psi_B^{\mu\nu},\chi^{II\mu}_B,\chi^{I\mu\nu}_+,\xi,\eta
\]
\[
Q=q,H^I_q
\]
\[
\psi_Q=\psi_q,\chi^I_q
\]
\[
I=\int \dfx( 
\qd(i\theta +m)q+\psiqd\psiq
)
\]
\begin{equation}
\Omega=\dim of~H's.
\end{equation}

In a general computation of the path integral of TFT,
it is sufficient 
to keep only quadratic terms for the transverse degrees
and compute the one-loop approximations which 
give a result exactly
\cite{TFT}.
Now let us see that in each branch, what is transverse degrees of freedom.
Picking a Riemannian metric $g $ ,
we rescale $g\to t g $ and take $t\to \infty $ limit.
In branch 1), the gauge symmetry is unbroken 
and the matter fields decouple as the transverse degrees of freedom.
In branch 2), the gauge symmetry is broken down to $U(1) $
and the hypermultiplet reduces to one of its color.
The suppressed color degrees of freedom for hypermultiplet
and the components of the $N=4$ vector multiplet 
which do not belong to the Cartan subalgebra part become
the transverse degrees of freedom.

On the other hand, the path integrals for the non-transverse degrees
should be computed exactly.
These path integrals correspond to the path integral
of Vafa-Witten theory in branch 1)
and the path integral of 
$U(1) $ monopole theory
and $U(1)B_+,c$  theory
in branch 2).

We will use the notation $<O>_{m,c,k} $ 
for the VEV evaluated 
in the massive UBTQCD 
for a given $spin^c $ 
and instanton number $k$.

\paragraph{result of branch 1)}~\\

In this branch, the degrees of freedom 
for the hypermultiplet become the transverse degrees of freedom.
One can decompose the lagrangian  (\ref{3.10}) into two parts

\begin{equation}
{\cal L }\approx {\cal L }^{V-W}(1)+{\cal L }^t(1),
\end{equation}
where the Vafa-Witten part

\begin{eqnarray}
{\cal L}^{V-W}(1)
&=&
-H^{I\mu\nu a}_+\{
H^{Ia}_{+\mu\nu}-(
F_{+\mu\nu}^a -i[B_{+\mu\rho},B_{+\nu\sigma}]^ag^{\rho\sigma}
 -i[B_{+\mu\nu},c]^a
)
\}\nonum
\\
& &
-\chi^{I\mu\nu a}_+\{
-i[\chi^I_{+\mu\nu},\theta]^a
+2D_\mu\psi_\nu^a
-2i[B_{+\mu\rho},\psi_{B\nu\sigma}]^ag^{\rho\sigma}
-i[\psi_{B\mu\nu},c]^a
-i[B_{+\mu\nu},\xi]^a
\}\nonum
\\
& &
-H^{II\rho a}_B\{
H^{IIa}_{B\rho}-(
-2D^\mu B_{+\mu\rho}^a
-D_\rho c^a
)
\}\nonum
\\
& &
-\chi^{II\rho a}_B\{
-i[\chi^{II}_{B\rho},\theta]^a
-2D^\mu\psi_{B\mu\rho}^a
-2i[\psi^\mu,B_{+\mu\rho}]^a
-D_\rho\xi^a
-i[\psi_\rho,c]^a
\}\nonum
\\
& &
-\{
[\theta,{\bar{\theta}}]^a[{\bar{\theta}},\theta]^a
-[c,\theta]^a[c,{\bar{\theta}}]^a
+[B^{\mu\nu}_+,{\bar{\theta}}]^a[B_{+\mu\nu},\theta]^a
\}+D_\mu{\bar{\theta}}^aD^\mu\theta^a
\nonum
\\
& &
+i[\theta,\eta]^a\eta^a
+i\xi^a[\xi,{\bar{\theta}}]^a
+i\xi^a[c,\eta]^a
+i[\psi^{\mu\nu}_B,{\bar{\theta}}]^a\psi_{B\mu\nu}^a
\nonum
\\
& &
+i[B^{\mu\nu}_+,\eta]^a\psi_{B\mu\nu}^a
+D_\mu\eta^a\psi^{\mu a}
+i[\psi_\mu,{\bar{\theta}}]^a\psi^{\mu a}
\nonum
\\
& &
\end{eqnarray}
and a quadratic lagrangian due to the transverse degrees

\begin{eqnarray}
{\cal L}^t(1)
&=&
-H^{I\dagger\alpha}_q\{
H^I_{q\alpha}-
s_\alpha
\}\nonum
\\
& &
-\chi^{I\dagger}_q
{\Sla D}\psi_q
\nonum
\\
& &+(h.c.~~ above~~ two~~ lines)
\nonum
\\
& &
-2\qd \mb m q +2\psiqd \mb \psi_q-2\chiqd m\chiq
\nonum
\\
&=&
-2\vert \Hq+\cdots\vert^2-2m\vert \chiq+\cdots\vert^2
\nonum
\\
& &
-\frac{1}{2}\qd
(
{\Sla D}^\dagger{\Sla D}+4m\mb
)
q
+\frac{1}{2m}\psiqd
(
{\Sla D}^\dagger{\Sla D}+4m\mb
)
\psiq
\label{3.291}
\end{eqnarray}

One can rewrite the path integral (\ref{3.29})  in this branch as

\begin{eqnarray}
<e^I>_{m,c,k}(1)
&=&
\underbrace{
\frac{1}{Vol{\cal G}(2\pi)^{\Omega^\prime}}
\int
\cD W\cD\psi_W
e^{-S^{V-W}(1)}
}_{\equiv Z_{m,c,k}^{V-W}(1)   }
\cdot
\underbrace{
\frac{1}{(2\pi)^{\Omega^{\prime\prime}}}
\int
\cD Q^\dagger\cD\psi^\dagger_Q \cD Q \cD \psi_Q
e^{-S^t(1)+I(1)}
}_{\equiv Z^t_{m,c,k}(1)},
\nonum
\\
& &
\end{eqnarray}
where
\[
h^2 S^{V-W}(1)=\int\dfx\sqrt g {\cal L}^{V-W}(1)
\]
\[
h^2 S^t(1)=\int\dfx\sqrt g {\cal L}^t(1)
\]
\[
I(1)=\int \dfx\sqrt g( 
\qd mq+\psiqd\psiq
)
\]
\[
\Omega^\prime=\dim of~adjoint~H's
\]
\begin{equation}
\Omega^{\prime\prime}=\dim of~fundamental~H's.
\end{equation}

For the Vafa-Witten part $ Z_{m,c,k}^{V-W}(1)  $,
we completely follow 
Vafa-Witten\cite{vafa-witten}.
Thus we have
\begin{equation}
Z_{m,c,k}^{V-W}(1) \doteq \chi_k,
\end{equation}
where $ \chi_k$ stands for the Euler number of 
instanton moduli space with instanton number $k $
and $\doteq$ means equality under keeping the vanishing theorem
 as shown in Vafa-Witten.
Note that the existence of the vanishing theorem in the previous section
is unknown, but, in this case, we have some examples to which we apply
the vanishing theorem \cite{vafa-witten}.  
When the vanishing theorem is not applicable,
we denote this part as $Z_{m,c,k}^{V-W}(1)$ itself.
We discuss the problem of compactification of moduli space later.

For the transverse part $Z^t_{m,c,k}(1)$, we first perform
$\Hq,\chiq $ integral and get
\begin{equation}
\frac{1}{(2\pi)^{\Omega^{\prime\prime}}}
\frac
{
[\det(-2m)]_{(\chiqd,\chiq)}
}
{
[\det(-\frac{1}{\pi})]_{(\Hqd,\Hq)}
}
=[\det (\frac{m}{2\pi} )]_{\Gamma^-}
=(\frac{m}{2\pi})^{dim(\Gamma^-_{\lambda>0}\oplus Ker({\sla D}^\dagger) )}.
\label{3.37}
\end{equation}

Second we perform 
$q,\psiq $ integral for zero and non-zero mode respectively and get

\begin{equation}
\frac
{
[\det
(
-\frac{{\sla {D}}^2}{2m}
)]_{(\psiqd,\psiq)_{non~0}}
}
{
[\det
(
-\frac{{\sla {D}}^2}{4\pi}
)]_{(\qd,q)_{non~0}}
}
\cdot
\frac
{
[\det
(
-1
)]_{(\psiqd,\psiq)_0}
}
{
[\det
(
-\frac{m}{2\pi}
)]_{(\qd,q)_0}
}
=(\frac{2\pi}{m})^{dim(\Gamma^+_{\lambda>0}\oplus Ker({\sla D})}.
\label{3.38}
\end{equation}
Note that this expression is not exact,
but is sufficient to get the right contribution (see Appendix).

Collecting  (\ref{3.37})  and  (\ref{3.38}), one can get

\begin{equation}
Z^t_{m,c,k}(1)=
(\frac{m}{2\pi})^{dim(\Gamma^-_{\lambda>0}\oplus Ker({\sla D}^\dagger) )}
(\frac{2\pi}{m})^{dim(\Gamma^+_{\lambda>0}\oplus Ker({\sla D})}
=(\frac{2\pi}{m})^{index({\sla D}^E_c)}.
\end{equation}

Finally for $<e^{\hat v}>_{m,c,k}(1)$, one can get

\begin{equation}
<e^{\hat v}>_{m,c,k}(1)=Z_{m,c,k}^{V-W}(1)
\cdot Z^t_{m,c,k}(1) 
=Z_{m,c,k}^{V-W}(1)\cdot(\frac{2\pi}{m})^{index({\sla D}^E_c)}
\doteq\chi_k \cdot(\frac{2\pi}{m})^{index({\sla D}^E_c)},
\end{equation}
where $\doteq $ stands for results in the vanishing theorem case.

\paragraph{result of branch 2)}~\\

In this branch, the gauge symmetry is broken down to $U(1) $.
The components of any field which do not belong to 
the Cartan subalgebra part become the transverse variables.
That is the $\pm $ components of adjoint fields
, i.e.  $T_\pm=T_1 \pm i T_2 $. 
And the components of the hypermultiplet with the suppressed 
color index become the transverse variable.
One can decompose  the lagrangian  (\ref{3.10})  into two parts

\begin{equation}
{\cal L}\approx  {\cal L}^{U(1)}(2)+{\cal L}^t(2),
\end{equation}
where ${\cal L}^{U(1)}(2) $ is the lagrangian of 
$U(1) $ UBTQCD,
and ${\cal L}^t(2)$ is  the quadratic lagrangian due to
the transverse degrees.

$U(1) $ part ${\cal L}^{U(1)}(2) $ can be further decomposed into 
two parts
\begin{equation}
  {\cal L}^{U(1)}(2)={\cal L}^{U(1)}_{mono}(2)+{\cal L}^{U(1)}_{B_+,c}(2),
\end{equation}

\begin{eqnarray}
{\cal L}^{U(1)}_{mono}(2)
&=&
-H^{I3\mu\nu}_+\{
H^{I3}_{+\mu\nu}-(
F_{+\mu\nu}^3+\half\qd_1\sigbmnd q_1
)
\}\nonum
\\
& &
-\chi^{I3\mu\nu}_+\{
2\nabla_\mu\psi_\nu^3
+\half{\psiqd}_1{\bar{\sigma}}_{\mu\nu}q_1
+\half {\qd}_1{\bar{\sigma}}_{\mu\nu}{\psiq}_1
\}\nonum
\\
& &
-{\Hqd}_1\{
{\Hq}_1-
{\Sla D}q_1
\}\nonum
\\
& &
-{\chiqd}_1
{\Sla D}{\psiq}_1
\nonum
\\
& &+(h.c.~~ above~~ two~~ lines)
\nonum
\\
& &
+\partial_\mu{\bar{\theta}}^3\partial^\mu\theta^3
+\partial_\mu\eta^3\psi^{3\mu}
\nonum
\\
& &
+2(
-i\half {\qd}_1{\bar{\theta}}^3-{\qd}_1{\bar{m}}
)
(i\half \theta^3 q_1+mq_1)
\nonum
\\
& &
+2{\psiqd}_1(i\half{\bar{\theta}}^3+{\bar{m}}){\psiq}_1
-2{\chiqd}_1(i\half\theta^3+m){\chiq}_1
\nonum
\\
& &
-(
-i\half {\qd}_1\eta^3-{\qd}_1\eta_m
){\psiq}_1
+{\psiqd}_1(
i\half\eta^3 q_1+\eta_m q_1
),\label{3.42}
\end{eqnarray}
and
\begin{equation}
  \label{2.352}
{\cal L}^{U(1)}_{B_+,c}(2)=
-H^{II3\rho}_B\{
H^{II3}_{B\rho}-(
-2\nabla^\mu B_{+\mu\rho}^3
-\partial_\rho c^3
)
\}
-\chi^{II3\rho}_B\{
-2\nabla^\mu\psi_{B\mu\rho}^3
-\partial_\rho\xi^3
\},
\end{equation}
where the first part ${\cal L}^{U(1)}_{mono}(2) $ is $U(1) $ monopole theory,
and the second part ${\cal L}^{U(1)}_{B_+,c}(2) $ is $U(1)~B_+,c $  theory.

The quadratic lagrangian due to
the transverse degrees ${\cal L}^t(2) $ is

\begin{eqnarray}
{\cal L}^t(2)
&= &
-4\vert H^{I+}_{+\mu\nu}+\cdots\vert^2
-8m\vert\chi^{I+}_{+\mu\nu}+\cdots\vert^2
+16m^2\vert \theb^++\cdots\vert^2
-8m\vert\eta^++\cdots\vert^2
\nonum
\\
& &
-A^+_\mu
\{
(D^{3+*}D^{3+})^{\mu\nu}+(D^3D^{3*})^{\mu\nu}
-{\tilde B}^{3\mu\rho}_+{\tilde B}^{3\nu}_{+~\rho}
-\half{\tilde q}^\dagger_1 \sigb^\mu\sigma^\nu{\tilde q}_1
+(-({\tilde c}^3)^2+16m\mb)\gmnu
\}
A^-_\nu
\nonum
\\
& &
+\frac{1}{2m}\psi^+_\mu
\{
(D^{3+*}D^{3+})^{\mu\nu}+(D^3D^{3*})^{\mu\nu}
-{\tilde B}^{3\mu\rho}_+{\tilde B}^{3\nu}_{+~\rho}
-\half{\tilde q}^\dagger_1 \sigb^\mu\sigma^\nu{\tilde q}_1
+( -({\tilde c}^3)^2  +16m\mb)\gmnu
\}
\psi^-_\nu
\nonum
\\
& &
-4\vert H^{II+}_B+\cdots\vert^2-8m\vert\chi^{II+}_B+\cdots\vert^2
\nonum
\\
& &
-\Bmnd
\{
(D^{3+}D^{3+*})^{\mu\rho}g^{\nu\sigma}
-4{\tilde B}^{3\mu\nu}_+{\tilde B}^{3\rho\sigma}_+
+(  -({\tilde c}^3)^2   +16m\mb)g^{\mu\rho}g^{\nu\sigma}
\}
B_{+\rho\sigma}
\nonum
\\
& &
+\frac{1}{2m}\psi_{B\mu\nu}
\{
(D^{3+}D^{3+*})^{\mu\rho}g^{\nu\sigma}
-4{\tilde B}^{3\mu\nu}_+{\tilde B}^{3\rho\sigma}_+
+( -({\tilde c}^3)^2    +16m\mb)g^{\mu\rho}g^{\nu\sigma}
\}
\psi_{B\rho\sigma}
\nonum
\\
& &
-c^+
\{
D^{3*}D^3-{\tilde B}^{3\mu\nu}_+{\tilde B}^3_{+\mu\nu}
-({\tilde c}^3)^2     -16m\mb
\}
c^-
\nonum
\\
& &
+\frac{1}{2m}\xi^+
\{
D^{3*}D^3-{\tilde B}^{3\mu\nu}_+{\tilde B}^3_{+\mu\nu}
-({\tilde c}^3)^2 -16m\mb
\}
\xi^-
\nonum
\\
& &
-2\vert {\Hq}_2+\cdots\vert^2
-4m\vert {\chiq}_2+\cdots\vert^2
\nonum
\\
& &
-\half\qd_2
\{
{\Sla D}^{3\dagger}{\Sla D}^3
-2\sigbmnd{\tilde q}_1 {\tilde q}^\dagger_1\sigbmnu
+16m\mb
\}
q_2
\nonum
\\
& &
+\frac{1}{4m}{\psiqd}_2
\{
{\Sla D}^{3\dagger}{\Sla D}^3
+2\sigbmnd{\tilde q}_1 {\tilde q}^\dagger_1\sigbmnu
+16m\mb
\}
{\psiq}_2
\nonum
\\
& &
+(cross~~terms).
\label{3.411}
\end{eqnarray}

One can rewrite the path integral (\ref{3.29}) in this branch as

\begin{eqnarray}
<e^I>_{m,c,k}(2)
&=&
\underbrace{
\frac{1}{Vol{\cal G}^3(2\pi)^{\Omega^\prime}}
\int
\cD W^3\cD\psi_W^3\cD Q^\dagger_1\cD\psi^\dagger_{Q1} \cD Q_1 \cD \psi_{Q1}
e^{-S^{U(1)}(2)}
}_{\equiv Z_{m,c,k}^{U(1)}(2)}
\nonum
\\
& &
\cdot
\underbrace{
\frac{1}{Vol{\cal G}^\pm(2\pi)^{\Omega^{\prime\prime}}}
\int
\cD W^\pm\cD\psi_W^\pm\cD Q^\dagger_2
\cD\psi^\dagger_{Q2} \cD Q_2 \cD \psi_{Q2}
e^{-S^t(2)+I(2)}
}_{\equiv Z^t_{m,c,k}(2)}
\nonum
\\
& &
\end{eqnarray}
where
\[
h^2 S^{U(1)}(2)=\int\dfx\sqrt g{\cal L}^{U(1)}(2)
\]
\[
h^2 S^t(2)=\int\dfx\sqrt g{\cal L}^t(2)
\]
\[
I(2)=\int \dfx\sqrt g( 
\qd_2 2mq_2+{\psiqd}_2{\psiq}_2
)
\]
\[
\Omega^\prime=\dim of~H's~ of~ non-transverse~ degrees
\]
\begin{equation}
\Omega^{\prime\prime}=\dim of~H's~ of~ transverse~ degrees.
\end{equation}

For $U(1) $ monopole part,
we have
\begin{equation}
\label{3.441}
Z_{mono}^{U(1)} = \frac{1}{Vol{\cal G}^3(2\pi)^{\Omega^{\prime\prime\prime}}}
\int
\cD W^3_A\cD\psi_{W^3_A}
\cD Q^\dagger_1\cD\psi^\dagger_{Q1} \cD Q_1 \cD \psi_{Q1}
e^{-S^{U(1)}_{mono}(2)},
\end{equation}
where
\[
W^3_A={\Amu}^3,H_{+\mu\nu}^3,\theta^3,{\theb}^3,q_1,{\Hq}_1
\]
\[
\psi_{W_A}^3=\psi_A^{\mu 3},\chi_{+\mu\nu }^3,\eta^3,{\psiq}_1,{\chiq}_1
\]
\[
h^2 S^{U(1)}_{mono}(2)=\int\dfx\sqrt g {\cal L}^{U(1)}_{mono}(2)
\]
\begin{equation}
\Omega^{\prime\prime\prime}=\dim of~H's~ of~ U(1)~S-W~ part.
\end{equation}
For this part we follow H-P-P \cite{H-P-P}.
In a simple type manifold 
we only need to consider 
the zero-dimensional moduli space of
the Seiberg-Witten monopoles (we call ${\cal M}(x) $).
Here we denote $spin^c $ structure $c^\prime $ 
that we have already mentioned in subsection \ref{sec:3-3}
by $2x$
if $c^\prime$ satisfies the condition of 
the zero-dimensional moduli space
($\dim {\cal M}(c^\prime)
=\frac{c^\prime\cdot c^\prime}{4}-\frac{2\chi+3\sigma}{4}=0$),
and we call this $spin^c $ structure $ x$  Seiberg-Witten basic class.
The moduli space ${\cal M}(x) $ consists of 
a finite set of points.
First for the contributions of
the zero-dimensional moduli space ${\cal M}(x) $,
we have 
\begin{equation}
\label{3.451}
{\cal N}n_x, 
\end{equation}
where ${\cal N} $ is the standard renormalization
due to the local operators constructed from
metric and depends only on $\chi $ and $\sigma $ \cite{S-W}.
$n_x $ is the sum of the number of points
counted with a sign and is called the Seiberg-Witten invariants.
For the total contribution to $U(1) $ monopole part (\ref{3.441}),
we have to sum (\ref{3.451}) with all 
basic classes $x $ and get
\begin{equation}
Z^{U(1)}_{mono}={\cal N}\sum_x n_x, 
\end{equation}

For $U(1)~B_+,c $ part we have
\begin{equation}
\label{3.45}
Z_{B_+}^{U(1)}
=\frac{1}{(2\pi)^{\Omega^{\prime\prime\prime\prime}}}
\int{\cal D}W^3_{B_+,c}{\cal D}\psi_{W_{B_+,c}}^3
e^{-S^1_{B_+,c}(2)},
\end{equation}
where
\[
W^3_{B_+,c}={\Bmnu}^3,H^{\mu 3}_B,c^3
\]
\[
\psi_{W_{B_+,c}}^3=\psi_B^{\mu\nu 3},\chi^{II\mu 3}_B,\xi^3
\]
\begin{equation}
  \label{2.351}
h^2 S^{U(1)}_{B_+,c}(2)=\int\dfx\sqrt g {\cal L}^{U(1)~B_+,c}(2)
\end{equation}
\[
\Omega^{\prime\prime\prime\prime}=\dim of~H's~ of~ U(1)~B_+,c~ part.
\]
$Z_{B_+}^{U(1)} $ is the partition function
of the cohomological field theory 
with the fixed point
\begin{equation}
\label{3.48}
\nabla^\mu {\Bmnd}^3 =0,\partial_\nu c^3=0.  
\end{equation}
This partition function is sum of 
the $\pm 1$ when there are only isolated solutions 
as usual. 
The condition that the $Z_{B_+}^{U(1)} $ is non-zero 
is that the dimensions of the moduli space 
of the 0 section defined by (\ref{3.48})
becomes zero.
In fact  the virtual dimension of this moduli space is calculated
to be
\begin{equation}
\Delta=index(d^{*+}+d)=\frac{1}{2}(\chi+\sigma),  
\end{equation}
where  $\chi $ and $\sigma$ are Euler number and signature
of $X$ respectively.
Thus $\Delta=0 $ is a condition that we get non-trivial results.
We discuss this point later.

Finally we get
\begin{equation}
Z_{m,c,k}^{U(1)}(2)
={\cal N} Z_{B_+}^{U(1)}  \sum_x n_x.
\end{equation}

Now we evaluate the transverse integral $Z^t_{m,c,k}(2) $.
Following H-P-P\cite{H-P-P}, we choose 
a unitary gauge in which
\begin{equation}
\theta_\pm=0, 
\end{equation}
where
\begin{equation}
\theta=\theta^3 T^3+\theta^+ T^++\theta^- T^-.
\end{equation}
In this gauge $\theta $ has values on the maximal torus
(Cartan sub-algebra).
By following the standard Faddev-Povov gauge fixing procedure,
we introduce a new nilpotent BRST operator $\delta $
with the algebra

\begin{equation}
\delta\theta_\pm=\pm iC_\pm \theta_3,\delta C_\pm=0,
\delta \theta_3=0,\delta {\bar C}_\pm=b_\pm,\delta b_\pm=0,
\end{equation}
where $C_\pm $ and ${\bar C}_\pm $ are anti-commuting
ghosts and anti-ghosts, respectively,
and $b_\pm $ are commuting auxiliary fields.
The action for gauge fixing terms reads
\begin{eqnarray}
S_{m,gauge}(2)&=&
\delta\frac{1}{mh^2}\int d^4 x\sqrt g(\theta_+{\bar C}_-+{\bar C}_+\theta_-)
\nonum
\\
&=&
 \frac{1}{mh^2} \int d^4 x\sqrt g\{
\theta_+b_-+b_+\theta_-
+iC_+\theta_3{\bar C}_-+i{\bar C}_+\theta_3C_-
\}\nonum
\\
&=&
\frac{1}{mh^2} \int d^4 x\sqrt g\{
\theta_+b_-+b_+\theta_-
-C_+2m{\bar C}_--{\bar C}_+2mC_-
\}.
\end{eqnarray}
  From the second line to the third line,
we take weak coupling limit and replace $\theta^3 $
with $2im $.
Note that this action has ghost number $0 $.

Now consider the transverse part involving adjoint fields.
We perform
$b_\pm,C_\pm,{\bar C}_\pm ,{\bar \theta}^\pm,\eta^\pm $ integral
and get
\begin{equation}
[\det(im)]_{\Omega^0}^{\half 2}[\det(-2)]_{\Omega^0}^{\half 2}
[\det(\frac{16m^2}{\pi})]^{-\half}_{\Omega^0}[\det(-8m)]_{\Omega^0}^\half
=[\det(2\pi m)]_{\Omega^0}^\half
=(2\pi m)^{\half dim(\Omega^0_{\lambda>0}\oplus Ker(D^3))} .
\end{equation}
\begin{itemize}
\item  $H_+^\pm,\chi_+^\pm $ integral

\begin{equation}
[\det(2\pi m)]_{\Omega^{2+}}^\half
=(2\pi m)^{\half dim(\Omega^{2+}_{\lambda>0}\oplus Ker(D^{3+*}))}.
\end{equation}

\item
$H_B^\pm,\chi_B^\pm $ integral

\begin{equation}
[\det(2\pi m)]_{\Omega^{1}}^\half
=(2\pi m)^{\half dim(\Omega^{1}_{\lambda>0}\oplus Ker(D^{3+}+D^{3*}))}.
\end{equation}
\item
$A^\pm,\psi^\pm $ integral for non-zero mode

\begin{equation}
\frac
{
[\det
(
-\frac{D^{3}D^{3*}+D^{3+*}D^{3+}}{m}
)]_{\psi^\pm_{non~0}}^\half
}
{
[\det
(
-\frac{D^{3}D^{3*}+D^{3+*}D^{3+} }{2\pi}
)]_{A^\pm_{non~0}}^\half
}
=
[\det(\frac{2\pi}{m})]_{\Omega^1_{non~0}}^\half
=
(\frac{2\pi}{m})^{\half dim(\Omega^1_{\lambda>0})}.
\end{equation}

\item
$B^\pm_+,\psi^\pm_B $ integral for non-zero mode

\begin{equation}
\frac
{
[\det
(
-\frac{D^{3+}D^{3+*}}{m}
)]_{{\chi^\pm_B}_{non~0}}^\half
}
{
[\det
(
-\frac{D^{3+}D^{3+*} }{2\pi}
)]_{{B^\pm_+}_{non~0}}^\half
}
=
[\det(\frac{2\pi}{m})]_{\Omega^{2+}_{non~0}}^\half
=
(\frac{2\pi}{m})^{\half dim(\Omega^{2+}_{\lambda>0})}
\end{equation}

\item
$c^\pm,\xi^\pm $ integral for non-zero mode

\begin{equation}
\frac
{
[\det
(
-\frac{D^{3*}D^{3}}{m}
)]_{\xi^\pm_{non~0}}^\half
}
{
[\det
(
-\frac{D^{3*}D^{3} }{2\pi}
)]_{c^\pm_{non~0}}^\half
}
=
[\det(\frac{2\pi}{m})]_{\Omega^0_{non~0}}^\half
=
(\frac{2\pi}{m})^{\half dim(\Omega^{0}_{\lambda>0})}.
\end{equation}
\end{itemize}

Now we collect all the contributions of the adjoint transverse part
and get

\begin{eqnarray}
\frac{1}{(2\pi)^{\Omega^{\prime\prime}_{ado}}}
(2\pi m)^{\half dim(\Omega^0_{\lambda>0}\oplus Ker(D^3))}
(2\pi m)^{\half dim(\Omega^{2+}_{\lambda>0}\oplus Ker(D^{3+*}))}
(2\pi m)^{\half dim(\Omega^{1}_{\lambda>0}\oplus Ker(D^{3+}+D^{3*}))}
\nonum
& &
\\
\cdot
(\frac{2\pi}{m})^{\half dim(\Omega^1_{\lambda>0})}
(\frac{2\pi}{m})^{\half dim(\Omega^{2+}_{\lambda>0})}
(\frac{2\pi}{m})^{\half dim(\Omega^{0}_{\lambda>0})}
&= &1
\nonum
\\
& &\label{3.59}
\end{eqnarray}

Remaining transverse integral is fundamental part.
First we perform
${\Hq}_2,{\chiq}_2 $ integral
and get

\begin{equation}
\frac{1}{(2\pi)^{\Omega^{\prime\prime}_{fun}}}
\frac
{
[\det(-4m)]_{(\chi_{q2}^\dagger,\chi_{q2})}
}
{
[\det(-\frac{1}{\pi})]_{(H_{q2}^\dagger,H_q^2)}
}
=[\det (\frac{m}{\pi})]_{\Gamma^-}
=(\frac{m}{\pi})^{dim(\Gamma^-_{\lambda>0}\oplus Ker(({\sla D}^3)^\dagger))}.
\label{3.60}
\end{equation}

Next we perform
$q_2,\psi_{q2}$ integral
for non-zero and zero mode respectively
and get
\begin{equation}
\frac
{
[\det
(
-\frac{{\sla D}^\dagger{\sla D} }{4m}
)]_{(\psi_{q2}^\dagger,{\psiq}_2)_{non~0}}
}
{
[\det
(
-\frac{{\sla D}^\dagger{\sla D}}{4\pi}
)]_{(q^\dagger_2,q_2)_{non~0}}
}
\frac
{
[\det
(
-1
)]_{(\psi_{q2}^\dagger,{\psiq}_2)_0}
}
{
[\det
(
-\frac{m}{\pi}
)]_{(q^\dagger_2,q_2)_0}
}
=
[\det(\frac{\pi}{m})]_{\Gamma^+}
=
(\frac{\pi}{m})^{dim(\Gamma^+_{\lambda>0}\oplus Ker((\sla {D}^3)))}.
\label{3.61}
\end{equation}

Collecting (\ref{3.60})  and (\ref{3.61}),  one can get
\begin{equation}
(\frac{m}{\pi})^{dim(\Gamma^-_{\lambda>0}\oplus Ker(({\sla D}^3)^\dagger))}
(\frac{\pi}{m})^{dim(\Gamma^+_{\lambda>0}\oplus Ker((\sla {D}^3)))}
=(\frac{\pi}{m})^{index({\sla D}^3)}
\label{3.62}
\end{equation}

  From (\ref{3.59})  and  (\ref{3.62})
we get
\begin{equation}
Z^t_{m,c,k}(2)=
(\frac{\pi}{m})^{index({\sla D}^3)}.
\label{3.671}
\end{equation}

Finally for 
$<e^I>_{m,c,k}(2)$
we get
\begin{equation}
<e^I>_{m,c,k}(2)=Z_{m,c,k}^{U(1)}(2)
\cdot
Z^2_{m,c,k}(2)
={\cal N}(\frac{\pi}{m})^{index({\sla D}^3)}
Z_{B_+} ^{U(1)} \sum_x n_x .
\end{equation}

\paragraph{synthesis}~\\

As we have already mentioned ,
$<e^I>_{m,c,k} $ itself is zero.
However from above two paragraphs each branch
has non-trivial contributions.
Thus we have finally 
\begin{eqnarray}
0&=&Z^{V-W}_{m,c,k} \cdot
\left(\frac{2\pi}{m}\right)^{index~ {\sla D}_c^E}
+{\cal N} Z_{B_+} \sum_x n_x  \cdot
\left(\frac{\pi}{m}\right)^{index~ {\sla D}^3}
\nonum
\\
&\doteq&
\chi_k \cdot
\left(\frac{2\pi}{m}\right)^{index~ {\sla D}_c^E}
+{\cal N} Z_{B_+} \sum_x n_x  \cdot
\left(\frac{\pi}{m}\right)^{index~ {\sla D}^3},
\label{3.66}
\end{eqnarray}
where the last expression is valid in the vanishing theorem case.

In general $index{\Sla D}^E_c $ is calculated to be
\begin{equation}
  index{\Sla D}^E_c=-k+\frac{rank(E)}{8}(c\cdot c -\sigma).
\end{equation}
In this case,
\begin{equation}
c \cdot \zeta = - index{\Sla D}^E_c + 2\Delta.
\end{equation}
The Dirac operator ${\Sla D}^3$ which 
operate on $q_2, \psi_{q_2}$, and so on
is necessary to be understood 
as the Dirac operator with the connection
given by $c-2\zeta$.
Then  
\begin{eqnarray}
  index{\Sla D}^3_c &=& 0+\frac{1}{8}
((c-2\zeta )\cdot (c-2\zeta )-\sigma )\\
 &=&\frac{1}{8}(c \cdot c -4k+4 index{\Sla D}^E_c
-8\Delta -\sigma) .
\end{eqnarray}
Thus we get a relation
\begin{equation}
  index{\Sla D}^3=index{\Sla D}^E_c-\Delta
\label{3.67}
\end{equation}

Inserting (\ref{3.67}) into (\ref{3.66}),
since 
$m $ is a free parameter,
we get non-trivial result 
only in the case $\Delta=0 $.
Remember that $\Delta $ is also the dimension 
of the moduli space 
of $U(1)~B_+,c $ theory. 
Thus the condition $\Delta=0 $ is consistent
with defining $Z^{U(1)}_{B_+} $   (\ref{3.45}).
$\Delta=0 $ is also consistent with 
geographic condition, for example
simple type condition($b^+_2>1 $), 
Furuta theory($b_2\geq \frac{5}{4}\vert \sigma\vert+2 $) and
$\frac{11}{8}$ conjecture($b_2\geq \frac{11}{8}\vert \sigma\vert $) \cite{geogra} \cite{furuta}.

Finally under the condition $\Delta=0 $,
from (\ref{3.66}) we have
\begin{equation}
\label{3.65}
  \chi_k\doteq Z^{V-W}_{m,c,k}=
-{\cal N} Z_{B_+} \sum_x n_x 
\left(\frac{1}{2}
\right)^{index {\sla D}^E_c}.
\end{equation}
Note that above $x$ satisfies that 
$x \cdot x = \frac{2 \chi + 3\sigma}{4}$ and 
$x=\frac{c+2\zeta}{2}$.

We think the Vafa-Witten partition as the sum of (\ref{3.65})
with weight $e^{\tau k} $, where $\tau$
is a parameter.
But the sum of this partition function
don not clarify modular invariance
since $\Delta =0$ is special case which
do not depend on the coupling $\tau$ in topological
twisted model \cite{vafa-witten} .
Additionally we do not assume duality, then
there is no guarantee that our partition
function has modular invariance and 
is same as Vafa-Witten's. 
We suppose that the difference come from 
compactification of the moduli space.
We do not use the duality relation and 
our model is not asymptotic-free theory. 
So, there is possibility
that compactification
in our theory is not the same as the 
one in Hilbert scheme.  
Thus we can describe the twisted $N=4$ Yang-Mills
 partition function that may not be the same as Vafa-Witten's 
partition function with 
Seiberg-Witten invariants.
Our expression is similar to Dijkgraaf\cite{DPS}.
The most significant difference is $\tau $ dependence.
Dijkgraaf's is $\tau $ dependent,
while ours is $\tau $ independent.
The reason why Dijkgraaf's partition function
depends on $\tau $ 
is that they treat the physical $N=4 $ Yang-Mills theory itself.
According to Labastida\cite{BTFT},
the $N=4 $ Yang-Mills theory depends on $\tau $.
On the other hand we treat UBTQCD,
which is the twisted $N=4 $ Yang-Mills theory
coupled with a fundamental hypermultiplet.
As we mention above, this difference 
may cause breaking the modular invariance.
In other words, our theory is not 
conformal invariant, and $\tau $
is not possible to be a good parameter. 
But our computation is done without
assumption like duality relation.
If there is difference  
we have to interpret the origin 
of the difference occurred from
 compactification \cite{furuta2}.

\section{Conclusion}
\hspace*{5mm}
We have studied the balanced topological QCD and 
its broken balance theory
and got relations of the partition function of twisted $N=4$
SU(2) Yang-Mills theory with the partition function
 of twisted abelian QCD.
This relation is understood in several ways.
For example, the sum of Euler number of instanton moduli space
, which is invariant under $ SL(2,Z) $ transformation, 
is described by Seiberg-Witten invariants
when $ \Delta = 0 $ and the vanishing theorem is valid.
In other cases there is no vanishing theorem like 
\S 5.4 in \cite{vafa-witten} ,
we got a similar but not the 
same formulas under the condition of 
$ \Delta =0 $.
There is no other reasons to understand the
 difference from the result of 
Vafa-Witten and Dijkgraaf et al.\cite{vafa-witten}
\cite{DPS} than the difference of compactification.
\\


Some problems are left for our future work.
When $ \Delta \neq 0 $, can we
obtain any similar non-trivial results without 
assumption of duality relation?
We may obtain them by simple reformation.
But it is difficult to expect that the partition
function have the nature of modular invariance
in naive reformation.    
We are interested in a connection 
with the duality 
and a compactification.
How can we obtain the modular invariant partition
function with no assumption of duality?
We have some hints of this question but
no answer.\\
As we saw in section 2, vanishing theorem of BTQCD is 
not studied in this paper.
If the theorem exists, we get the sum of Euler number 
of non-abelian monopole moduli space as the partition function
of the BTQCD.
It is an interesting work to investigate the nature of the 
partition function because the theory has the branches
that contain both Vafa-Witten theory and Seiberg-Witten theory.\\

{ \Large Acknowledgment}\\

We are grateful to H.Kanno for helpful suggestions and 
observations and a critical reading of the manuscript.
We also would like to thank M.Furuta and K.Ono 
for valuable discussion.
A.S. is supported by JSPS Research Fellowships for 
Young Scientists.

\appendix
\section{the BRS algebra and the BTQCD action}
\setcounter{equation}{0}
\renewcommand{\theequation}{\alph{section}.\arabic{equation}.\arabic{subeq}}
\addtocounter{subeq}{1}
\hspace*{5mm}
We give the BRS algebra and the lagrangian
of BTQCD explicitly in this appendix.

\subsection{Algebra} 
\hspace*{5mm}
$\delta_{\pm}$ transformations are given
as follows.
\begin{equation}
  \label{2.6.1}
\left\{
\begin{array}{c}
\delta_-A_\mu=\chi^{II}_\mu\\
\delta_-\chi^{II}_\mu=-\delta^{\bar{\theta}}_gA_\mu
=-D_\mu{\bar{\theta}}\\
\delta_-\psi_\mu=-\delta^c_gA_\mu-H^{II}_{B\mu}
=-D_\mu c-H^{II}_{B\mu} \\
\delta_-H^{II}_{B\mu}=-\delta^c_g\chi^{II}_{B\mu}
+\delta_+\delta^{\bar{\theta}}_gA_\mu
=-i[\chi^{II}_{B\mu},c]+\delta_+\delta^{\bar{\theta}}_gA_\mu
\end{array}
\right.
\end{equation}
\addtocounter{subeq}{1}
\addtocounter{equation}{-1}

\begin{equation}
 \left\{
\begin{array}{c}
\delta_-B_+^{\mu\nu}=\chi^{I\mu\nu}_+\\
\delta_-\chi^{I\mu\nu}_+=-\delta^{\bar{\theta}}_gB_+^{\mu\nu}
=-i[B^{\mu\nu}_+,{\bar{\theta}}]\\
\delta_-\psi_B^{\mu\nu}=-\delta^c_gB_+^{\mu\nu}-H^{I\mu\nu}_+
=-i[B^{\mu\nu}_+,c]-H^{I\mu\nu}_+\\
\delta_-H^{I\mu\nu}=-\delta^c_g\chi^{I\mu\nu}_+
+\delta_+\delta^{\bar{\theta}}_gB_+^{\mu\nu}
=-i[\chi^{I\mu\nu}_+,c]+\delta_+\delta^{\bar{\theta}}_gB_+^{\mu\nu}
\end{array}
\right.
\end{equation}
\addtocounter{subeq}{1}
\addtocounter{equation}{-1}

\begin{equation}
\left\{
\begin{array}{c}
\delta_-q^{\dot{\alpha}}=\chi^{II\dot{\alpha}}_B\\
\delta_-\chi^{II\dot{\alpha}}_B=\delta^{\bar{\theta}}_gq^{\dot{\alpha}}
=i{\bar{\theta}}q^{\dot{\alpha}}+{\bar{m}}q^{\dot{\alpha}}
\\
\delta_-\psi_q^{\dot{\alpha}}=\delta^c_gq^{\dot{\alpha}}-H^{II\dot{\alpha}}_B
=icq^{\dot{\alpha}}+m_cq^{\dot{\alpha}}-H^{II\dot{\alpha}}_B
\\
\delta_-H^{II\dot{\alpha}}_B=\delta^c_g\chi^{II\dot{\alpha}}_B
-\delta_+\delta^{\bar{\theta}}_gq^{\dot{\alpha}}
=ic\chi^{II\dot{\alpha}}_B+m_c
-\delta_+\delta^{\bar{\theta}}_gq^{\dot{\alpha}}
\end{array}
\right.
\end{equation}
\addtocounter{subeq}{1}
\addtocounter{equation}{-1}

\begin{equation}
\left\{
\begin{array}{c}
\delta_-B_\alpha=\chi^I_{q\alpha}\\
\delta_-\chi^I_{q\alpha}=\delta^{\bar{\theta}}_gB_\alpha
=i{\bar{\theta}}B_\alpha+{\bar{m}}B_\alpha
\\
\delta_-\psi_{B\alpha}=\delta^c_gB_\alpha-H^I_{q\alpha}
=icB_\alpha+m_cB_\alpha-H^I_{q\alpha}
\\
\delta_-H^I_{q\alpha}=\delta^c_g\chi^I_{q\alpha}
-\delta_+\delta^{\bar{\theta}}_gB_\alpha
=ic\chi^I_{q\alpha}+m_c\chi^I_{q\alpha}
-\delta_+\delta^{\bar{\theta}}_gB_\alpha 
\end{array}
\right. 
\end{equation}
\addtocounter{subeq}{1}
\addtocounter{equation}{-1}

\begin{equation}
\left\{
\begin{array}{c}
\delta_-q^\dagger_{\dot{\alpha}}=\chi^{II\dagger}_{B\dot{\alpha}}\\
\delta_-\chi^{II\dagger}_{B\dot{\alpha}}
=\delta^{\bar{\theta}}_gq^\dagger_{\dot{\alpha}}
=-iq^\dagger_{\dot{\alpha}}{\bar{\theta}}-{\bar{m}}q^\dagger_{\dot{\alpha}}
\\
\delta_-\psi_{q\dot{\alpha}}^\dagger
=\delta^c_gq^\dagger_{\dot{\alpha}}-H^{II\dagger}_{B\dot{\alpha}}
=-iq^\dagger_{\dot{\alpha}}c-m_cq^\dagger_{\dot{\alpha}}
-H^{II\dagger}_{B\dot{\alpha}}
\\
\delta_-H^{II\dagger}_{B\dot{\alpha}}
=\delta^c_g\chi^{II\dagger}_{B\dot{\alpha}}
-\delta_+\delta^{\bar{\theta}}_gq^\dagger_{\dot{\alpha}}
=-i\chi^{I\dagger}_{B\dot{\alpha}}c-m_c\chi^{I\dagger}_{B\dot{\alpha}}
-\delta_+\delta^{\bar{\theta}}_gq^\dagger_{\dot{\alpha}}
\end{array}
\right.
\end{equation}
\addtocounter{subeq}{1}
\addtocounter{equation}{-1}

\begin{equation}
\left\{
\begin{array}{c}
\delta_-B^{\dagger\alpha}=\chi^{I\dagger\alpha}_q  \\
\delta_-\chi^{I\dagger\alpha}_q=\delta^{\bar{\theta}}_gB^{\dagger\alpha}  
=-iB^{\dagger\alpha}{\bar{\theta}}-{\bar{m}}B^{\dagger\alpha}
\\
\delta_-\psi^{\dagger\alpha}_B=\delta^c_gB^{\dagger\alpha}
-H^{I\dagger\alpha}_q
=-iB^{\dagger\alpha}c-m_cB^{\dagger\alpha}-H^{I\dagger\alpha}_q
\\
\delta_-H^{I\dagger\alpha}_q
=\delta^c_g\chi^{I\dagger\alpha}_q
-\delta_+\delta^{\bar{\theta}}_gB^{\dagger\alpha}
=-i\chi^{I\dagger\alpha}_qc-m_c\chi^{I\dagger\alpha}_q
-\delta_+\delta^{\bar{\theta}}_gB^{\dagger\alpha}.
\end{array}
\right.
\end{equation}
\setcounter{subeq}{1}

$\delta_+$ transformations are given by

\begin{equation}
  \label{2.7.1}
\left\{
\begin{array}{c}
\delta_+A_\mu=\psi_\mu  
\\
\delta_+\psi_\mu=\delta^\theta_gA_\mu  
=D_\mu\theta
\\
\delta_+\chi_{B\mu}^{II}=H^{II}_{B\mu}
\\
\delta_+H^{II}_{B\mu}
=\delta^\theta_g\chi^{II}_{B\mu}
=i[\chi^{II}_{B\mu},\theta]
\end{array}
\right.
\end{equation}
\addtocounter{subeq}{1}
\addtocounter{equation}{-1}

\begin{equation}
\left\{
\begin{array}{c}
\delta_+B_+^{\mu\nu}=\psi_B^{\mu\nu}  \\
\delta_+\psi_B^{\mu\nu} =\delta^\theta_gB_+^{\mu\nu}  
=i[B^{\mu\nu}_+,\theta]
\\
\delta_+\chi^{I\mu\nu}_+=H^{I\mu\nu}_+
\\
\delta_+H^{I\mu\nu}_+
=\delta^\theta_g\chi^{I\mu\nu}_+
=i[\chi^{I\mu\nu}_+,\theta]
\end{array}
\right.
\end{equation}
\addtocounter{subeq}{1}
\addtocounter{equation}{-1}

\begin{equation}
\left\{
\begin{array}{c}
\delta_+q^{\dot{\alpha}}=\psi^{\dot{\alpha}}_q  \\
\delta_+\psi^{\dot{\alpha}}_q =-\delta^\theta_g q^{\dot{\alpha}}_q  
=-(i\theta q^{\dot{\alpha}}_q+mq^{\dot{\alpha}}_q)
\\
\delta_+\chi^{II\dot{\alpha}}_B=H^{II\dot{\alpha}}_B
\\
\delta_+H^{II\dot{\alpha}}_B
=-\delta^\theta_g\chi^{II\dot{\alpha}}_B
=-(i\theta\chi^{II\dot{\alpha}}_B+m\chi^{II\dot{\alpha}}_B)
\end{array}
\right.
\end{equation}
\addtocounter{subeq}{1}
\addtocounter{equation}{-1}

\begin{equation}
\left\{
\begin{array}{c}
\delta_+B_\alpha=\psi_{B\alpha}  \\
\delta_+\psi_{B\alpha}=-\delta^\theta_gB_\alpha  
=-(i\theta B_\alpha+mB_\alpha)
\\
\delta_+\chi^I_{q\alpha}=H^I_{q\alpha}
\\
\delta_+H^I_{q\alpha}
=-\delta^\theta_g\chi^I_{q\alpha}
=-(i\theta\chi^I_{q\alpha}+m\chi^I_{q\alpha})
\end{array}
\right.
\end{equation}
\addtocounter{subeq}{1}
\addtocounter{equation}{-1}

\begin{equation}
\left\{
\begin{array}{c}
\delta_+q^\dagger_{\dot{\alpha}}=\psi^\dagger_{q\dot{\alpha}}  \\
\delta_+\psi^\dagger_{q\dot{\alpha}}
=-\delta^\theta_gq^\dagger_{\dot{\alpha}}  
=-(-iq^\dagger_{\dot{\alpha}}\theta-mq^\dagger_{\dot{\alpha}})
\\
\delta_+\chi^{II\dagger}_{B\dot{\alpha}}=H^{II\dagger}_{B\dot{\alpha}}
\\
\delta_+H^{II\dagger}_{B\dot{\alpha}}
=-\delta^\theta_g\chi^{II\dagger}_{B\dot{\alpha}}
=-(-i\chi^{II\dagger}_{B\dot{\alpha}}\theta-m\chi^{II\dagger}_{B\dot{\alpha}})
\end{array}
\right.
\end{equation}
\addtocounter{subeq}{1}
\addtocounter{equation}{-1}

\begin{equation}
\left\{
\begin{array}{c}
\delta_+B^{\dagger\alpha}=\psi^{\dagger\alpha}_B  \\
\delta_+\psi^{\dagger\alpha}_B
=-\delta^\theta_gB^{\dagger\alpha}  
=-(-iB^{\dagger\alpha}\theta-mB^{\dagger\alpha})
\\
\delta_+\chi^{I\dagger\alpha}_q=H^{I\dagger\alpha}_q
\\
\delta_+H^{I\dagger\alpha}_q
=-\delta^\theta_g\chi^{I\dagger\alpha}_q
=-(-i\chi^{I\dagger\alpha}_q\theta-m\chi^{I\dagger\alpha}_q).
\end{array}
\right.
\end{equation}
\setcounter{subeq}{1}

Transformations for
$c,\theta,\bar{\theta},m,m_c,{\bar{m}}$
are given by

\begin{equation}
  \label{2.8}
\left\{
\begin{array}{cc}
\delta_+\theta=0  \\
\delta_-\theta=\xi &,\delta_+\xi=\delta^\theta_gc=i[c,\theta] \\
\delta_+c=\xi&,\delta_-\xi=\delta^{\bar{\theta}}_g\theta
=i[\theta,{\bar{\theta}}]  \\
\delta_-c=\eta &,\delta_+\eta=\delta^\theta_g{\bar{\theta}}
=i[{\bar{\theta}},\theta]\\
\delta_+{\bar{\theta}}=\eta&,\delta_-\eta=\delta^{\bar{\theta}}_g c
=i[c,{\bar{\theta}}]  \\
\delta_-{\bar{\theta}}=0
\end{array}
\right.
\end{equation}
\addtocounter{subeq}{1}
\addtocounter{equation}{-1}
\begin{equation}
\label{2.8.2}
\left\{
\begin{array}{cc}
\delta_+m=0  \\
\delta_-m=\xi_m &,\delta_+\xi_m=0 \\
\delta_+m_c=\xi_m&,\delta_-\xi_m=0\\
\delta_-m_c=\eta_m &,\delta_+\eta_m=0\\
\delta_+{\bar{m}}=\eta_m&,\delta_-\eta_m=0  \\
\delta_-{\bar{m}}=0.
\end{array}
\right.
\end{equation}

\renewcommand{\theequation}{\alph{section}.\arabic{equation}}

\subsection{action of BTQCD}
\hspace*{5mm}
We write down the lagrangian of BTQCD explicitly
in this paragraph.

$ \delta_- {\cal F} $ is given as 

\begin{eqnarray}
\delta_-{\cal F}&=&
\delta_-(B^{\mu\nu a}_+s_{+\mu\nu}^a)
-\delta_-(\chi^{I\mu\nu a}_+\psi_{B\mu\nu}^a)
-\delta_-(\chi^{IIa}_{B\mu}\psi^{\mu a})
+\delta_-(-i\frac{1}{3}{\Bmnu}^a[B_{+\mu\rho},B_{+\nu\sigma}]^ag^{\rho\sigma})
\nonum
\\
& &
+\delta_-(B^{\dagger\alpha}s_\alpha)
-\delta_-(\chi^{I\dagger\alpha}_q\psi_{B\alpha})
-\delta_-(\chi^{II\dagger}_{B\dot{\alpha}}\psi^{\dot{\alpha}}_q)
\nonum
\\
& &
+\delta_-(s^{\dagger\alpha}B_\alpha)
+\delta_-(\psi^{\dagger\alpha}_B\chi^I_{q\alpha})
+\delta_-(\psi^\dagger_{q\dot{\alpha}}\chi^{II\dot{\alpha}}_B)
\nonum
\\
& &
+\delta_-(\xi^a\eta^a)
\label{2.11}
\\
&=&
-\chi^{I\mu\nu a}_+\{
H^{Ia}_{+\mu\nu }-(
s_{+\mu\nu}^a-i[B_{+\mu\rho},B_{+\nu\sigma}]^ag^{\rho\sigma} 
-i[B_{+\mu\nu},c]^a
)
\}
\nonum
\\
& &
-\chi^{II\rho a}\{
H^{IIa}_{B\rho}-(
-2D^\mu B_{+\mu\rho}^a
+iB^\dagger \sigma_\rho T^a q
-iq^\dagger{\bar{\sigma}_\rho}T^a B
-D_\rho c^a
)
\}
\nonum
\\
& &
-\chi^{I\dagger\alpha}_q\{
H^I_{q\alpha}-(
s_\alpha+icB_\alpha+m_cB_\alpha
)
\}
\nonum
\\
& &
-\chi^{II\dagger}_{B\dot{\alpha}}\{
H^{II\dot{\alpha}}_B-(
-(\Sla{D}^\dagger B)^{\dot{\alpha}}
+({\bar{\sigma}}^{\mu\nu}B_{+\mu\nu}  q)^{\dot{\alpha}}
+icq^{\dot{\alpha}}+m_cq^{\dot{\alpha}}
)
\}
\nonum
\\
& &
-\{
H^{I\dagger\alpha}_q-(
s^{\dagger\alpha}
-iB^{\dagger\alpha}c-m_cB^{\dagger\alpha}
)
\}\chi^I_{q\alpha}
\nonum
\\
& &
-\{
H^{II\dagger}_{B\dot{\alpha}}-(
-({\Sla{D}}^\dagger B)^\dagger_{\dot{\alpha}}
+(q^\dagger   B_{+\mu\nu} {\bar{\sigma}}^{\mu\nu})_{\dot{\alpha}}
-iq^\dagger_{\dot{\alpha}}c-m_cq^\dagger_{\dot{\alpha}}
)
\}\chi^{II\dot{\alpha}}_B
\nonum
\\
& &
+\{ i[\theta,{\bar{\theta}}]^a\eta^a
-i\xi^a[c,{\bar{\theta}}]^a \}
+i[B^{\mu\nu}_+,{\bar{\theta}}]^a\psi_{B\mu\nu}^a
+D_\mu{\bar{\theta}}^a\psi^{\mu a}
\nonum
\\
& &
-(
-iB^{\dagger\alpha}{\bar{\theta}}-{\bar{m}}B^{\dagger\alpha}
)\psi_{B\alpha}
-(
-iq^\dagger_{\dot{\alpha}}{\bar{\theta}}-{\bar{m}}q^\dagger_{\dot{\alpha}}
)\psi^{\dot{\alpha}}_q
\nonum
\\
& &
-\psi^{\dagger\alpha}_B(
i{\bar{\theta}}B_\alpha+{\bar{m}}B_\alpha
)
-\psi^\dagger_{q\dot{\alpha}}(
i{\bar{\theta}}q^{\dot{\alpha}}+{\bar{m}}q^{\dot{\alpha}}
),\label{2.12}
\end{eqnarray}

The full lagrangian is given as 

\begin{eqnarray}
{\cal L}^{full}&=&\delta_+\delta_-{\cal F}
\label{2.15}
\\
&=&
-H^{I\mu\nu a}_+\{
H^{Ia}_{+\mu\nu}-(
s_{+\mu\nu}^a-i[B_{+\mu\rho},B_{+\nu\sigma}]^ag^{\rho\sigma}
-i[B_{+\mu\nu},c]^a
)
\}
\nonum
\\
& &
-\chi^{I\mu\nu a}_+\{
-i[\chi^I_{+\mu\nu},\theta]^a
+2D_\mu\psi_\nu^a
+\psi^\dagger_q{\bar{\sigma}}_{\mu\nu}T^a q
+q^\dagger{\bar{\sigma}}_{\mu\nu}T^a \psi_q
-2i[B_{+\mu\rho},\psi_{B\nu\sigma}]^ag^{\rho\sigma}
\nonum
\\
& &
-i[\psi_{B\mu\nu},c]^a
-i[B_{+\mu\nu},\xi]^a
\}
\nonum
\\
& &
-H^{II\rho a}_B\{
H^{IIa}_{B\rho}-(
-2D^\mu B_{+\mu\rho }^a
+iB^\dagger \sigma_\rho T^a q-iq^\dagger{\bar{\sigma}_\rho}T^a B
-D_\rho c^a
)
\}
\nonum
\\
& &
-\chi^{II\rho a}_B\{
-i[\chi^{II}_{B\rho},\theta]^a
-2D^\mu\psi_{B\mu\rho}^a
-2i[\psi^\mu,B_{+\mu\rho}]^a
\nonum
\\
& &
+i\psi^\dagger_B\sigma_\rho T^aq
+iB^\dagger\sigma_\rho T^a \psi_q
-i\psi^\dagger_q{\bar{\sigma}}_\rho T^a B
-iq^\dagger{\bar{\sigma}}_\rho T^a \psi_B
-D_\rho\xi^a
-i[\psi_\rho,c]^a
\}
\nonum
\\
& &
-H^{I\dagger\alpha}_q\{
H^I_{q\alpha}-(
s_\alpha+icB_\alpha+m_cB_\alpha
)
\}
\nonum
\\
& &
-\chi^{I\dagger}_q\{
{\Sla D}\psi_q
+\sigma_\rho i\psi^\rho q
+i\xi B+ic\psi_B
+\xi_mB+m_c\psi_B
\}
\nonum
\\
& &+(h.c.~~ above~~ two~~ lines)
\nonum
\\
& &
-H^{II\dagger}_{B\dot{\alpha}}\{
H^{II\dot{\alpha}}_B-(
-(\Sla{D}^\dagger B)^{\dot{\alpha}}
+({\bar{\sigma}}^{\mu\nu} B_{+\mu\nu}   q)^{\dot{\alpha}}
+icq^{\dot{\alpha}}+m_cq^{\dot{\alpha}}
)
\}
\nonum
\\
& &
-\chi^{II\dagger}_B\{
-{\Sla D}^\dagger \psi_B
-{\bar{\sigma}}_\rho i\psi^\rho B
+({\bar{\sigma}}^{\mu\nu}\psi_{B\mu\nu}q)
+({\bar{\sigma}}^{\mu\nu}B_{+\mu\nu}\psi_q)
+i\xi q+ic\psi_q
+\xi_mq+m_c\psi_q
\}
\nonum
\\
& &+(h.c.~~ above~~ two~~ lines)
\nonum
\\
& &
-\{
[\theta,{\bar{\theta}}]^a[{\bar{\theta}},\theta]^a
-[c,\theta]^a[c,{\bar{\theta}}]^a
+[B^{\mu\nu}_+,{\bar{\theta}}]^a[B_{+\mu\nu},\theta]^a
\}+D_\mu{\bar{\theta}}^aD^\mu\theta^a
\nonum
\\
& &
+i[\theta,\eta]^a\eta^a
+i\xi^a[\xi,{\bar{\theta}}]^a
+i\xi^a[c,\eta]^a
+i[\psi^{\mu\nu}_B,{\bar{\theta}}]^a\psi_{B\mu\nu}^a
\nonum
\\
& &
+i[B^{\mu\nu}_+,\eta]^a\psi_{B\mu\nu}^a
+D_\mu\eta^a\psi^{\mu a}
+i[\psi_\mu,{\bar{\theta}}]^a\psi^{\mu a}
\nonum
\\
& &
+(
-iq^\dagger{\bar{\theta}}-q^\dagger{\bar{m}}
)
(i\theta q+mq)
+(
-iq^\dagger\theta-q^\dagger m
)
(i{\bar{\theta}}q+{\bar{m}}q)
\nonum
\\
& &
+2\psi^\dagger_q(i{\bar{\theta}}+{\bar{m}})\psi_q
-2\chi^{I\dagger}_q(i\theta+m)\chi^I_q
-(
-iq^\dagger\eta-q^\dagger\eta_m
)\psi_q
+\psi^\dagger_q(
i\eta q+\eta_m q
)
\nonum
\\
& &
+(
-iB^\dagger{\bar{\theta}}-B^\dagger{\bar{m}}
)
(i\theta B+mB)
+(
-iB^\dagger\theta-B^\dagger m
)
(i{\bar{\theta}}B+{\bar{m}}B)
\nonum
\\
& &
+2\psi^\dagger_B(i{\bar{\theta}}+{\bar{m}})\psi_B
-2\chi^{II\dagger}_B(i\theta+m)\chi^{II}_B
-(
-iB^\dagger\eta-B^\dagger\eta_m
)\psi_B
+\psi^\dagger_B(
i\eta B+\eta_m B
).
\nonum
\\
& &
\label{full}
\end{eqnarray}

\section{the path integral of the transverse part}
\setcounter{equation}{0}
\hspace*{5mm}
As we have mentioned in the first part of section\ref{sec:3-4},
the path integral in \ref{sec:3-4} is not exact,
but it amounts to the right result 
that we will derive in this section.
In computation,
we take the weak coupling limit.
When we replace the non-transverse fields
with the fixed point values,
we denote $Y_{non-trans} $ by ${\tilde Y}_{non-trans} $.
In particular
the fixed points of $\theta ,\theb $  
are given as $\theta=\theb=0 $ in branch 1)
and $\theta^3=2im, \theb=2i\mb $ in branch 2).
We also discuss the different treatment from 
the path integral in \ref{sec:3-4}
at the end of this section.
See \cite{sako}, too.
\subsection{branch 1) and its big matrix}
\hspace*{5mm}
In branch 1),
the path integral of the transverse part is

\begin{eqnarray}
 Z^t_{m,c,k}(1)
&=&
\frac{1}{(2\pi)^{\Omega^{\prime\prime}}}
\int
\cD Q^\dagger\cD\psi^\dagger_Q \cD Q \cD \psi_Q
e^{-S^t(1)+I(1)}
\label{a.1}
\end{eqnarray}
where
\[
h^2 S^t(1)=\int\dfx\sqrt g {\cal L}^t(1)
\]
\[
I(1)=\int \dfx\sqrt g( 
\qd mq+\psiqd\psiq
)
\]
\begin{equation}
\Omega^{\prime\prime}=\dim of~fundamental~H's.
\end{equation}

For ${\cal L}^t(1)$(\ref{3.291}),
we denote
\begin{eqnarray}
  {\cal L}^t(1)
&=&
-2\vert \Hq+\cdots\vert^2-2m\vert \chiq+\cdots\vert^2
\nonum
\\
& &
-\frac{1}{2}\qd
M^b(1)
q
+\frac{1}{2m}\psiqd
M^f(1)
\psiq,
\end{eqnarray}
where
\begin{equation}
M^b(1)=M^f(1)={\Sla D}^\dagger{\Sla D}+4m\mb.
\label{a.101}
\end{equation}
In general $M^b(M^f) $ is matrix 
and is not necessarily diagonalized.
$M^b$ and $M^f$ may not be the same as we will see soon. 
We call $M^b(1)$  big matrix of branch 1).

Before computing (\ref{a.1}),
we briefly review the notion of $index{\Sla D} $.

One can decompose $q\in \Gamma^+ , \Hq\in \Gamma^-$ into
\[
{\Sla D}^\dagger{\Sla D} q^\lambda = \lambda q^\lambda
\] 
\begin{equation}
{\Sla D}{\Sla D}^\dagger {\Hq}^\lambda = \lambda {\Hq}^\lambda.
\end{equation}
These decomposition is called spectra decomposition.
Note that if $\lambda> 0 $ then $q^\lambda $ and ${\Hq}^\lambda$
are isomorphic.
However if $\lambda=0$ then $q^\lambda $ and ${\Hq}^\lambda$
are not isomorphic.
$index{\Sla D}  $ measures the difference 
between $\Gamma_{\lambda=0}^+ $ and $\Gamma_{\lambda=0}^- $,
and is defined as
\begin{eqnarray}
  index{\Sla D}&=&\dim \Gamma^+_{\lambda=0}-\dim \Gamma^-_{\lambda=0}
\nonum
\\
&=&
\dim Ker{\Sla D}-\dim Ker {\Sla D}^\dagger,
\label{a.2}
\end{eqnarray}
where we denote $\Gamma^+_{\lambda=0}=Ker{\Sla D},
\Gamma^+_{\lambda=0}=Ker {\Sla D}^\dagger$.

In computing (\ref{a.1}),
(\ref{a.2}) emerges 
when non-kinetic part and off-diagonal part of $M $ are 
able to be ignored 
(in this branch simply $m\mb $ terms in (\ref{a.101})).
This process is achieved by diagonalization and field
redefinition.
Then we get the expression (\ref{a.1})
as $index{\Sla D} $.
Conversely it is enough to get this expression
that we consider only kinetic diagonal part of $M $
in the path integral.

Now we perform the path integral of the transverse part of branch 1) explicitly.

First for 
$\Hq,\chiq $ integral, 
\begin{equation}
\frac{1}{(2\pi)^{\Omega^{\prime\prime}}}
\frac
{
[\det(-2m)]_{(\chiqd,\chiq)}
}
{
[\det(-\frac{1}{\pi})]_{(\Hqd,\Hq)}
}
=[\det (\frac{m}{2\pi} )]_{\Gamma^-}
=(\frac{m}{2\pi})^{dim(\Gamma^-_{\lambda>0}\oplus Ker({\sla D}^\dagger) )}.
\label{a.201}
\end{equation}
Note that the transformation at the second equality 
is necessary to derive $index{\Sla D} $.

For  
$q,\psiq $ integral for non-zero mode, 
\begin{equation}
\frac
{
[\det
(
-\frac{{\sla {D}}^2}{2m}
)]_{(\psiqd,\psiq)_{non~0}}
}
{
[\det
(
-\frac{{\sla {D}}^2}{4\pi}
)]_{(\qd,q)_{non~0}}
}
=(\frac{2\pi}{m})^{dim(\Gamma^+_{\lambda>0})}.
\label{a.3}
\end{equation}

For  
$q,\psiq $ integral for zero mode,
we consider that the integrant of this path integral
comes only from observable $I(1) $
and we get  
\begin{equation}
\frac
{
[\det
(
-1
)]_{(\psiqd,\psiq)_0}
}
{
[\det
(
-\frac{m}{2\pi}
)]_{(\qd,q)_0}
}
=(\frac{2\pi}{m})^{\dim Ker({\sla D})}.
\label{a.4}
\end{equation}

  From (\ref{a.3}) and (\ref{a.4})
\begin{equation}
  (\frac{2\pi}{m})^{\dim (\Gamma^+_{\lambda>0} \oplus Ker({\sla D}))}.
\label{a.5}
\end{equation}

Collecting (\ref{a.201}) and (\ref{a.5}),
\begin{equation}
Z^t_{m,c,k}=
  (\frac{m}{2\pi})^{dim(\Gamma^-_{\lambda>0}\oplus Ker({\sla D}^\dagger) )}
(\frac{2\pi}{m})^{\dim (\Gamma^+_{\lambda>0} \oplus Ker({\sla D}))}
=(\frac{2\pi}{m})^{index{\sla D}}.
\label{a.6}
\end{equation}
Note that $ dim(\Gamma^+_{\lambda>0})$ and
 $ dim(\Gamma^-_{\lambda>0})$  cancel each other.

\subsection{branch 2) and its big matrix}
\hspace*{5mm}
In branch 2),
the path integral of the transverse part is

\begin{eqnarray}
Z^t_{m,c,k}(2)
&=&
\frac{1}{Vol{\cal G}^\pm(2\pi)^{\Omega^{\prime\prime}}}
\int
\cD W^\pm\cD\psi_W^\pm\cD Q^\dagger_2
\cD\psi^\dagger_{Q2} \cD Q_2 \cD \psi_{Q2}
e^{-S^t(2)+I(2)},
\nonum
\\
& &
\label{a.7}
\end{eqnarray}
where
\[
h^2 S^t(2)=\int\dfx\sqrt g{\cal L}^t(2)
\]
\[
I(2)=\int \dfx\sqrt g( 
\qd_2 2mq_2+{\psiqd}_2{\psiq}_2
)
\]
\begin{equation}
\Omega^{\prime\prime}=\dim of~H's~ of~ transverse~ degrees.
\end{equation}

For ${\cal L}^t(2) $ (\ref{3.411}),
we denote

\begin{eqnarray}
{\cal L}^t(2)
&= &
-4\vert H^{I+}_{+\mu\nu}+\cdots\vert^2
-8m\vert\chi^{I+}_{+\mu\nu}+\cdots\vert^2
+16m^2\vert \theb^++\cdots\vert^2
-8m\vert\eta^++\cdots\vert^2
\nonum
\\
& &
-4\vert H^{II+}_B+\cdots\vert^2-8m\vert\chi^{II+}_B+\cdots\vert^2
\nonum
\\
& &
-2\vert {\Hq}_2+\cdots\vert^2
-4m\vert {\chiq}_2+\cdots\vert^2
\nonum
\\
& &
-Y^T M^b(2) Y+\frac{1}{2m} \Psi_Y^T M^f(2) \Psi_Y,
\label{a.8}
\end{eqnarray}
where $Y^T, \Psi_Y^T $ are raw vectors,
\begin{equation}
  Y^T=(A^+_\mu,{B_{+\nu\rho}}^+,c^+,{\qd}_2)
\end{equation}
\begin{equation}
 \Psi_Y^T=(\psi^+_\mu,{\psi_{B\nu\rho}}^+,\xi^+,{\psiqd}_2),
\end{equation}
and $Y, \Psi_Y $ are column vectors,
\begin{equation}
  Y=(A^-_\sigma,{B_{+\gamma\delta}}^-,c^-,q_2)
\end{equation}
\begin{equation}
 \Psi_Y=(\psi^-_\sigma,{\psi_{B\gamma\delta}}^-,\xi^-,{\psiq}_2).
\end{equation}

To derive the result (\ref{3.671}) from (\ref{a.7}) (\ref{a.8}),
we can neglect the non-kinetic terms and off-diagonal part of $M(2) $
(we will give explicitly later).
There is the contribution from 
Faddeev-Popov determinant of $\theta^\pm=0 $ gauge
and it is possible to discard the path integral of $Y^\pm $ 
for zero mode 
according to the balanced structure of adjoint fields.

In this remaining subsection,
we concentrate on giving $M(2) $ explicitly.
$M^b(2)(M^f(2)) $ can be decomposed into
\begin{equation}
M^b(2)=\left(
  \begin{array}{cc}
M^b_{AA}&M^b_{Aq}\\
M^b_{\qd A}&M^b_{\qd q}
  \end{array}
\right).
\end{equation}
We denote matrix element of $M^b(2)$ 
(or $M^b_{AA},M^b_{Aq},M^b_{\qd A},M^b_{\qd q}$)
by 
$\{M^b(2)\}^{A^{+\mu}B_+^{-\gamma\delta}},
\{M^b_{Aq} \}^{A^{+\mu}q} $ etc.

\paragraph{diagonal part of $M^b(2)(M^f(2)) $}

\begin{eqnarray}
\{M^b_{AA}\}^{A^{+\mu}A^{-\sigma}}&=&
\{M^f_{AA}\}^{A^{+\mu}A^{-\sigma}}\nonum
\\
&=&
(D^{3+*}D^{3+})^{\mu\sigma}+(D^3D^{3*})^{\mu\sigma}
-{\tilde B}^{3\mu\rho}_+{\tilde B}^{3\sigma}_{+~\rho}
-\half{\tilde q}^\dagger_1 \sigb^\mu\sigma^\sigma{\tilde q}_1
+( -({\tilde c}^3)^2  +16m\mb)g^{\mu\sigma}
\nonum
\\
& &
\label{a.20}
\end{eqnarray}
\begin{eqnarray}
\{M^b_{AA}\}^{B_+^{+\nu\rho}B_+^{-\gamma\delta}}&=&
\{M^f_{AA}\}^{B_+^{+\nu\rho}B_+^{-\gamma\delta}}\nonum
\\
&=&
(D^{3+}D^{3+*})^{\nu\rho}g^{\gamma\delta}
-4{\tilde B}^{3\nu\rho}_+{\tilde B}^{3\gamma\delta}_+
+(  -({\tilde c}^3)^2   +16m\mb)g^{\nu\rho}g^{\gamma\delta}
\nonum
\\
\end{eqnarray}

\begin{eqnarray}
\{M^b_{AA}\}^{c^+ c^-}&=&
\{M^f_{AA}\}^{c^+ c^-}\nonum
\\
&=&
D^{3*}D^3-{\tilde B}^{3\mu\nu}_+{\tilde B}^3_{+\mu\nu}
-({\tilde c}^3)^2     -16m\mb
\end{eqnarray}

\begin{eqnarray}
  \{M^b_{\qd q}\}^{\qd q}&=&
 {\Sla D}^{3\dagger}{\Sla D}^3
-2\sigbmnd{\tilde q}_1 {\tilde q}^\dagger_1\sigbmnu
+16m\mb
\label{a.23}
\end{eqnarray}

\begin{eqnarray}
  \{M^f_{\qd q}\}^{\qd q}&=&
{\Sla D}^{3\dagger}{\Sla D}^3
+2\sigbmnd{\tilde q}_1 {\tilde q}^\dagger_1\sigbmnu
+16m\mb
\label{a.24}
\end{eqnarray}
Note that $\{M^b_{\qd q}\}^{\qd q}$ and $\{M^f_{\qd q}\}^{\qd q}$
are different.

\paragraph{off-diagonal part of $M^b_{AA}(M^f_{AA})$}

\begin{eqnarray}
 \{M^b_{AA}\}^{A^{+\mu}B_+^{-\rho\sigma}}
&=&
\{M^b_{AA}\}^{A^{+\mu}B_+^{-\rho\sigma}}
\nonum
\\
&=&i({\overleftarrow {D^{3+}}})_\nu {\tilde c}^3 g^{\mu\rho}g^{\nu\sigma}
-2i{\tilde B}_+^{3\mu\rho}(D^{3+*})^\rho
-i({\overleftarrow {D^{3*}} })^\mu {\tilde B}_+^{3\rho\sigma}
\nonum
\\
& &
-2i({\overleftarrow {D^{3+}}})_\nu {\tilde B}_+^{3\mu\rho}g^{\nu\sigma}
\end{eqnarray}

\begin{eqnarray}
 \{M^b_{AA}\}^{A^{+\mu}c^-}
&=&
\{M^b_{AA}\}^{A^{+\mu}c^-}
\nonum
\\
&=&-i({\overleftarrow {D^{3+}}})_\nu {\tilde B}_+^{3\mu\nu}
-2i{\tilde B}_+^{3\mu\rho}(D^3)_\rho
+i({\overleftarrow {D^{3*}} })^\mu {\tilde c}^3
\end{eqnarray}

\begin{eqnarray}
 \{M^b_{AA}\}^{B_+^{+\nu\rho}c^-}
&=&
\{M^b_{AA}\}^{B_+^{+\nu\rho}c^-}
\nonum
\\
&=&({\overleftarrow {D^{3+*}}})^\nu (D^3)^\rho
-2{\tilde c}^3{\tilde B}_+^{3\nu\rho}
\end{eqnarray}

\paragraph{off-diagonal part of $M^b_{Aq}(M^f_{Aq})$}~\\

Here using
\begin{equation}
Y^\pm=\half(Y^1\mp i Y^2),  
\end{equation}
we denote $\{M^b_{Aq}\}^{Y^1q}, \{M^b_{Aq}\}^{Y^2q}$ and
$\{M^f_{Aq}\}^{Y^1q}, \{M^f_{Aq}\}^{Y^2q}$ 
instead of 
$\{M^b_{Aq}\}^{Y^+q}, \{M^f_{Aq}\}^{Y^+q}$.
The reason why we cannot denote 
$\{M^b_{Aq}\}^{Y^+q}, \{M^f_{Aq}\}^{Y^+q}$
is that there are terms 
$(D_\mu^3 A_\nu^+-D_\nu^3 A_\mu^+)_+{\qd}_2\sigbmnu {\tilde q}_1 $
and
$i({\Sla D}^3 q_2)^\dagger {\Sla A^-}{\tilde q}_1$
 exist simultaneously in ${\cal L}^t(2) $ (\ref{a.8})
, for example.

\begin{equation}
\{M^b_{Aq} \}^{A^{1\mu }q}
=-\half ({\overleftarrow {D^{3+}}})^\nu {\tilde q}^\dagger_1\sigbmnd
-i\frac{1}{4}  {\tilde q}^\dagger_1{\bar \sigma}_\mu {\Sla D}^3
-i\frac{1}{4}({\overleftarrow {D^{3+}}})^\mu  {\tilde q}^\dagger_1
\label{a.29}
\end{equation}

\begin{equation}
\{M^b_{Aq} \}^{A^{2\mu }q}
=-i\half ({\overleftarrow {D^{3+}}})^\nu {\tilde q}^\dagger_1\sigbmnd
-\frac{1}{4}  {\tilde q}^\dagger_1{\bar \sigma}_\mu {\Sla D}^3
-\frac{1}{4}({\overleftarrow {D^{3+}}})^\mu  {\tilde q}^\dagger_1
\end{equation}

\begin{equation}
\{M^f_{Aq} \}^{A^{1\mu }q}
=\half ({\overleftarrow {D^{3+}}})^\nu {\tilde q}^\dagger_1\sigbmnd
-i\frac{1}{4}  {\tilde q}^\dagger_1{\bar \sigma}_\mu {\Sla D}^3
-i\frac{1}{4}({\overleftarrow {D^{3+}}})^\mu  {\tilde q}^\dagger_1
\end{equation}

\begin{equation}
\{M^b_{Aq} \}^{A^{2\mu }q}
=i\half ({\overleftarrow {D^{3+}}})^\nu {\tilde q}^\dagger_1\sigbmnd
-\frac{1}{4}  {\tilde q}^\dagger_1{\bar \sigma}_\mu {\Sla D}^3
-\frac{1}{4}({\overleftarrow {D^{3+}}})^\mu  {\tilde q}^\dagger_1
\end{equation}

\begin{equation}
  \{M^b_{Aq} \}^{B^{1\nu\rho }q}
=-i\half {\tilde c}^3{\tilde q}_1^\dagger {\bar \sigma}^{\nu\rho}
+\frac{1}{4}{\tilde B}^{3\nu\rho}{\tilde q}_1^\dagger
+i{\tilde B}_{+\mu}^{3~~\rho}\sigbmnu {\tilde q}_1^\dagger
\end{equation}

\begin{equation}
  \{M^b_{Aq} \}^{B^{2\nu\rho }q}
=\half {\tilde c}^3{\tilde q}_1^\dagger {\bar \sigma}^{\nu\rho}
-i\frac{1}{4}{\tilde B}^{3\nu\rho}{\tilde q}_1^\dagger
-{\tilde B}_{+\mu}^{3~~\rho}\sigbmnu {\tilde q}_1^\dagger
\end{equation}

\begin{equation}
  \{M^f_{Aq} \}^{B^{1\nu\rho }q}
=i\half {\tilde c}^3{\tilde q}_1^\dagger {\bar \sigma}^{\nu\rho}
+\frac{1}{4}{\tilde B}^{3\nu\rho}{\tilde q}_1^\dagger
-i{\tilde B}_{+\mu}^{3~~\rho}\sigbmnu {\tilde q}_1^\dagger
\end{equation}

\begin{equation}
  \{M^f_{Aq} \}^{B^{2\nu\rho }q}
=-\half {\tilde c}^3{\tilde q}_1^\dagger {\bar \sigma}^{\nu\rho}
-i\frac{1}{4}{\tilde B}^{3\nu\rho}{\tilde q}_1^\dagger
+{\tilde B}_{+\mu}^{3~~\rho}\sigbmnu {\tilde q}_1^\dagger
\end{equation}

\begin{equation}
  \{M^b_{Aq} \}^{c^1q}
=i\half {\tilde B}_+^{3\mu\nu}\sigbmnd {\tilde q}_1^\dagger
-\frac{1}{4}{\tilde c}^3{\tilde q}_1^\dagger
\end{equation}

\begin{equation}
  \{M^b_{Aq} \}^{c^2q}
=-\half {\tilde B}_+^{3\mu\nu}\sigbmnd {\tilde q}_1^\dagger
+i\frac{1}{4}{\tilde c}^3{\tilde q}_1^\dagger
\end{equation}

\begin{equation}
  \{M^f_{Aq} \}^{c^1q}
=-i\half {\tilde B}_+^{3\mu\nu}\sigbmnd {\tilde q}_1^\dagger
-\frac{1}{4}{\tilde c}^3{\tilde q}_1^\dagger
\end{equation}

\begin{equation}
  \{M^f_{Aq} \}^{c^2q}
=\half {\tilde B}_+^{3\mu\nu}\sigbmnd {\tilde q}_1^\dagger
+i\frac{1}{4}{\tilde c}^3{\tilde q}_1^\dagger
\label{a.40}
\end{equation}
For (\ref{a.29})$\sim$(\ref{a.40}),
one can fine the relation
\begin{equation}
\left(
  \begin{array}{c}
\{M^f_{Aq} \}^{Y^1q}\\
\{M^f_{Aq} \}^{Y^2q}
  \end{array}
\right)=
\left(
  \begin{array}{cc}
0&i\\
-i&0
  \end{array}
\right)
  \left(
  \begin{array}{c}
\{M^b_{Aq} \}^{Y^1q}\\
\{M^b_{Aq} \}^{Y^2q}
  \end{array}
\right).
\label{a.41}
\end{equation}

Using above explicit matrix elements (\ref{a.20})$\sim$(\ref{a.40}),
we can perform the path integral (\ref{a.7}) directly,
instead of neglecting non-kinetic off-diagonal part of $M^b(2)(M^f(2)) $.
Then we have a crucial obstacle from the difference
between (\ref{a.23}) and (\ref{a.24}),
while the obstacle from (\ref{a.29})$\sim$(\ref{a.40}) is resolved
by the relation (\ref{a.41}).
This obstacle tells us 
that the contributions from (\ref{a.23}) and (\ref{a.24})
is not $1 $
and that
the result (\ref{3.671})
is effective up to order of square of $\tilde{q}_1$.
(In fact this problem does not appear 
when we treat adjoint matter instead of fundamental matter.
Thus we think that this problem comes from 
the choice of the representation of matter fields.)
However
the contributions from (\ref{a.23}) and (\ref{a.24})
becomes $1 $ in ${\tilde q}_1\to 0 $ limit after path 
integration.
Thus we estimate that  the contributions from (\ref{a.23}) and (\ref{a.24})
to be  $1 $ in the case that the result $Z^t_{m,c,k}(2) $ (\ref{3.671})
is topological.
This is why it is enough to estimate the path integral
with the indexes that 
the only kinetic terms in diagonal block 
are counted from the big matrices in section 3.

\end{document}